\documentclass[a4paper,fleqn,usenatbib]{mnras}
\pdfoutput=1
\usepackage[T1]{fontenc}
\usepackage{ae,aecompl}
\usepackage{amsmath}
\usepackage{amssymb}
\usepackage{txfonts}

\usepackage{graphicx}

\usepackage[usenames, dvipsnames]{color}
\usepackage{rotating}

\usepackage{soul}

\def\aj{AJ}%
\def\apj{ApJ}%
\def\apjl{ApJ}%
\def\apjs{ApJS}%
\def\ao{Appl.\ Opt.}%
\def\aap{A\&A}%
\def\jcap{JCAP}%
\def\mnras{MNRAS}%
\def\prd{Phys.\ Rev.\ D}%
\def\rmxaa{Rev. Mexicana Astron. Astrofis.}%
\def\nat{Nature}%
\def\bain{Bull.\ Astron.\ Inst.\ Netherlands}%
\def\physrep{Phys.\ Rep.}%
\def\repprogphys{Rep.\ Prog.\ Phys.}%
\def\ltsima{$\; \buildrel < \over \sim \;$}
\def\simlt{\lower.5ex\hbox{\ltsima}}   
\def\gtsima{$\; \buildrel > \over \sim \;$}
\def\simgt{\lower.5ex\hbox{\gtsima}}

\def \MultiNest{{\sc MultiNest\,}}
\def \Msun{{\rm M}_\odot}

\newcommand*\ruleline[1]{\par\noindent\raisebox{.8ex}{\makebox[0.55\linewidth]{\hrulefill\hspace{1ex}\raisebox{-.8ex}{#1}\hspace{1ex}\hrulefill}}}
\newcommand{\sd}{\mathit{SD}}

\title{A non-parametric method for measuring the local dark matter density}
\author[H. Silverwood et al.]{H. Silverwood$^{1}$, S. Sivertsson$^{1,2}$, P. Steger$^{3}$, J. I. Read$^{4}$, G. Bertone$^{1}$\\
$^1$ GRAPPA, University of Amsterdam, Science Park 904, 1098 XH Amsterdam, The Netherlands\\
$^2$ The Oskar Klein Centre for Cosmoparticle Physics, Department of Physics, Stockholm University, AlbaNova, SE-106 91 Stockholm, Sweden\\
$^3$ Institute for Astronomy, Department of Physics, ETH Z\"urich, Wolfgang-Pauli-Strasse 27, CH-8093 Z\"urich, Switzerland\\
$^4$ Department of Physics, University of Surrey, Guildford, GU2 7XH, Surrey, UK\\
}

\date{Accepted XXX. Received YYY; in original form ZZZ}
\pubyear{2016}

\begin{document}
\label{firstpage}
\pagerange{\pageref{firstpage}--\pageref{lastpage}}
\maketitle

\begin{abstract}
We present a new method for determining the local dark matter density using kinematic data for a population of tracer stars. The Jeans equation in the $z$-direction is integrated to yield an equation that gives the velocity dispersion as a function of the total mass density, tracer density, and the `tilt' term that describes the coupling of vertical and radial motions. We then fit a dark matter mass profile to tracer density and velocity dispersion data to derive credible regions on the vertical dark matter density profile. Our method avoids numerical differentiation, leading to lower numerical noise, and is able to deal with the tilt term while remaining one dimensional. In this study we present the method and perform initial tests on idealised mock data. We also demonstrate the importance of dealing with the tilt term for tracers that sample $\simgt 1$\,kpc above the disc plane. If ignored, this results in a systematic underestimation of the dark matter density.  
\end{abstract}


\section{Introduction}\label{sec:introduction}

Dark matter (DM) is an elusive component of the cosmos. Its existence has long been inferred from its gravitational interactions with ordinary matter on scales ranging from dwarf galaxies and galactic rotation curves to lensing galaxies, galaxy clusters, and the Universe as a whole (for reviews see e.g. \citealt{1993PNAS...90.4814R, Jungman96, Bergstrom00, Bertone05, BertoneBook, 2010RPPh...73h6901M}). However its exact nature remains unknown. 

Understanding the distribution of DM in cosmological structures, and in particular in the Milky Way, is of great importance to testing the standard $\Lambda$-Cold Dark Matter ($\Lambda$-CDM) cosmological model; to make predictions -- and  allow the interpretation -- of experiments that seek to detect DM; and to probe more exotic models for DM \citep[e.g.][]{Read:2014qva}. Here we will focus our attention on the local DM density -- i.e. the average density of DM over a small volume (typically of size $\sim$ 100pc - 1000pc) centred around the Sun. The local DM density is a vital input for so-called {\it direct} DM experiments, based on the detection of the recoil energy of nuclei struck by DM particles streaming through underground detectors, as the expected event rate is proportional to the local DM density and the DM-nucleon cross section. In addition, the local DM density is important for indirect searches that look for neutrinos produced by the annihilation of DM particles trapped at the centre of Sun (see e.g. \citealt{Klasen:2015uma} and references therein for a recent update on direct and indirect searches). The results of these searches are also used in explorations of theoretical parameter space, such as those of Supersymmetry \citep[e.g.][]{2006PhRvD..73a5013A, Buchmueller:2013rsa,Strege:2014ija,Roszkowski:2014iqa, 2016JCAP...04..037B}. 

Methods of determining the local DM density can be divided into two categories: those utilising measurements of stars in a small volume around the Sun \citep[e.g.][]{Kapteyn:1922zz, 1922MNRAS..82..122J, 1932BAN.....6..249O, 1984ApJ...287..926B, 1989MNRAS.239..605K, 1991ApJ...367L...9K, 1998A&A...329..920C, Garbari:2012ff,  2012ApJ...756...89B, 2012ApJ...746..181S, 2013ApJ...772..108Z, 2014A&A...571A..92B} and those utilising a host of dynamical tracers -- particularly the rotation curve -- to constrain more global mass models of the Milky Way \citep{Dehnen:1996fa,Weber:2009pt, CatenaUllio2010, Salucci:2010qr, 2011MNRAS.414.2446M, Nesti:2013uwa, 2014MNRAS.445.3133P, 2015NatPh..11..245I, Pato:2015tja, 2015arXiv150406324P}. Following the notation of \citet{Read:2014qva} we denote the local DM density derived from local measurements by $\rho_{\rm DM}$, and those extrapolated from rotation curves assuming a spherical DM halo as $\rho_{\rm DM, ext}$. With the advent of large survey data, local and global methods are beginning to converge \citep{2013ApJ...779..115B, 2014MNRAS.445.3133P}. 

The comparison of $\rho_{\rm DM}$ to $\rho_{\rm DM, ext}$ can provide insight into the shape of the Milky Way's DM halo, and thus into the formation of the galaxy \citep{Read:2014qva}. If $\rho_{\rm DM}$ from local measurements is smaller than that extrapolated from global measurements, $\rho_{\rm DM, ext}$, then this implies a prolate halo, while the opposite would imply a squashed, oblate halo. While the former is produced by DM only N-body simulations, the introduction of baryons into the simulations produces the latter \citep{1991ApJ...377..365K, Dubinski:1993df, 2008ApJ...681.1076D, Read:2009iv}. 
 
If $\rho_{\rm DM}$ is derived from tracers distributed vertically in the plane it is possible to probe not only the local DM density but also its vertical profile, $\rho_{\rm DM}(z)$. This could potentially reveal deviations from a spherical halo profile, which can form in a number of ways. First, the DM halo should respond to the formation of the baryonic disc by flattening into an oblate halo \citep{1991ApJ...377..365K, Dubinski:1993df, 2008ApJ...681.1076D, Read:2009iv}. Second, the accretion of subhalos after the formation of the baryonic disc can also lead to the formation of a `dark disc' that co-rotates with the baryonic disc \citep{1989AJ.....98.1554L, Read:2008fh, Read:2009iv, 2009ApJ...703.2275P, 2010JCAP...02..012L, 2014ApJ...784..161P}. Gravitationally these two features, an oblate halo and a dark disc, are indistinguishable and degenerate. One method to distinguish the two is to hunt for the chemically and kinematically distinct sub-halo stars that would accompany the accreted dark disc, but not the contracted DM halo. Recent Gaia-ESO observational data find no evidence for such stars \citep{2014MNRAS.444..515R, Ruchti:2015bja}, suggesting that the Milky Way has had a rather quiet past since its disc formed, and therefore does not host a significant accreted dark disc. If correct, then any gravitationally detected non-sphericity must imply a non-rotating locally oblate halo. An enhancement in $\rho_{\rm DM}$ over $\rho_{\rm DM, ext}$ is also a feature in some more exotic models, such as Partially Interacting Dark Matter (PIDM) \citep{Fan2013139}, and some modified theories of gravity \cite[e.g.][]{2005MNRAS.361..971R, 2007MNRAS.379..597N}. 

Measurement of the local DM density dates back almost 100 years \citep{Kapteyn:1922zz}. The history of this measurement is one of both increasingly precise data and decreasingly strong assumptions. As such, the error bars on $\rho_{\rm DM}$ have not always shrunk with time, as better data often allows one to discard previous assumptions. With the advent of Gaia data from 2016 onwards we will have access to high precision data on individual stars, and the primary uncertainty in the determination of $\rho_{\rm DM}$ will be systematic model uncertainties. Recent results for the local DM density include \cite{2014A&A...571A..92B}, who used RAVE data to derive a value of $\rho_{\rm DM} = 0.0143 \pm 0.0011 \Msun {\rm pc}^{-3} = 0.542\pm 0.042\, {\rm GeV} {\rm cm}^{-3}$, and \cite{2013ApJ...772..108Z} who used SDSS/SEGUE data to find $\rho_{\rm DM} = 0.0065 \pm 0.0023 \Msun {\rm pc}^{-3} = 0.25\pm 0.09\, {\rm GeV} {\rm cm}^{-3}$. Note that these two results do not overlap within their stated uncertainties. The significance of this discrepancy is hard to interpret though, as they each use different data sets and analysis techniques, and both methodologies rely on rather different assumptions. To make progress we should endeavour to minimise the number of assumptions made, and apply the same analysis techniques to both data sets.

In this paper we make progress towards the first goal of reducing model assumptions. To achieve this we introduce a one-dimensional Jeans analysis method to probe the local DM density using the vertical motions of tracer stars. We construct a representation of the tracer density $\nu$, and also allow for a DM density profile that is more complex than simply constant with height as previous local studies have assumed. Additionally, we deal with the so-called tilt term, which links radial and vertical motions of the tracer stars. This term is crucial to understand stellar motions at high-$z$, where the baryonic contribution falls off and DM becomes increasingly important. Using the vertical Jeans equation we calculate the velocity dispersion $\sigma_z(z)$ for each mass model, and then fit to data in $\nu$, $\sigma_z$, and $\sigma_{Rz}$ using \MultiNest ~\citep{Feroz:2007kg, Feroz:2008xx, 2013arXiv1306.2144F}.  We test this method on mock data sets, and explore the impact of tracer star sample size, the tilt term, and non-constant DM profiles. The mock data for this paper are `as good as it gets' -- we assume the population to have no observational biases and there to be no measurement error on individual stars -- allowing us to isolate the effects of sampling error and model uncertainties. This is similar to what we will have with Gaia data, where the measurement errors on stars will be small compared to sampling error. We will explore the effect of adding realistic Gaia uncertainties to our method in a separate work. 

We reiterate that this method is one dimensional, allowing us to keep assumptions to a minimum. However, we show that we are still able to deal with tilt and high-$z$ data, which usually requires a 2-dimensional method. The key disadvantages of our method are that we bin data and thus lose information, and that our strictly local method cannot take advantage of the `long lever arm' of a global model that would ensure, for example, that the DM density in radial slices close to the `local volume' is continuous and smooth. However the effect of these disadvantages is to overestimate the errors on $\rho_{\rm DM}$, which is acceptable as we aim for a conservative and robust estimate of $\rho_{\rm DM}$. 

The paper is organised as follows: in Section \ref{sec:method} we introduce our method, covering the Jeans equation based mathematical formalism, our treatment of the rotation curve and tilt term, our our descriptions of the elements of our mass and tracer density models.  We then outline our statistical analysis, introducing the framework for Bayesian parameter estimation and the \MultiNest nested sampling code. In Section \ref{sec:mock_data} we describe the array of mock data sets we use to test out methods. In Section \ref{sec:Results} we present and discuss the results of our analyses, before finally concluding the paper with Section \ref{sec:conclusions}. 

\section{Method} \label{sec:method}
The broad picture of our problem is this: we have quantities derived from the motions of the tracer stars, namely $\nu$, the tracer density, $\sigma_{z}$, the vertical velocity dispersion, and $\sigma_{R,z}$, the $(R,z)$ velocity dispersion. Then we have elements of the mass profile, one of which is unknown -- the DM density $\rho_{\rm DM}$, and the other which is known within a band of uncertainty -- the baryon density $\rho_{\rm baryon}$. In this section we first derive the equations to link these quantities, and then describe how each is modelled. 

\subsection{Deriving a general 1D Jeans method}\label{subsec:JeansEq}
A population of `tracer' stars moving vertically near the Sun obeys the collisionless Boltzmann equation: 
\begin{equation}
\frac{df}{dt} = \frac{\partial f}{\partial t} + \nabla_x f\cdot{\bf v} - \nabla_v f\cdot \nabla_x \Phi = 0
\label{eqn:colboltz}
\end{equation}
where $f({\bf x},{\bf v})$ is the stellar distribution function; ${\bf x}$ and ${\bf v}$ are the positions and velocities, respectively; and $\Phi$ is the gravitational potential.

If the system is in dynamic equilibrium (steady state), then we may neglect the partial time derivative of $f$. Thus for equilibrium tracers, if we know their phase space distribution function $f$, then we can directly measure the gravitational force from equation~(\ref{eqn:colboltz}). In practice, however, this is hard because $f$ is six-dimensional (even a million stars gives only 10 sample points per dimension) and we need to estimate the partial derivatives of $f$ to solve equation~(\ref{eqn:colboltz}). For this reason, it is common to integrate equation~(\ref{eqn:colboltz}) over velocity to obtain a set of moment equations: the Jeans equations \citep{Binney2008}. Adopting cylindrical polar coordinates $(R,\phi,z)$ and focussing on the $z$-Jeans equation, we have 
\begin{equation} 
\underbrace{\frac{1}{R\nu}\frac{\partial}{\partial R}\left(R\nu \sigma_{Rz}\right)}_{\text{`tilt' term}:\,\,\mathcal{T}} + \underbrace{\frac{1}{R\nu}\frac{\partial}{\partial \phi}\left(\nu \sigma_{\phi z}\right)}_{\text{`axial' term}:\,\,\mathcal{A}} + \frac{1}{\nu}\frac{d}{dz}\left(\nu\sigma_z^2\right) = \underbrace{-\frac{d\Phi}{dz}}_{K_z}
\label{eqn:zjeans}
\end{equation} 
where $\nu$ and $\sigma_z^2$ are the density and vertical velocity dispersion profile of a tracer population as a function of height $z$ above the disc plane, $\sigma_{Rz}$ is the cross term in the velocity dispersion tensor that couples radial and vertical motions, and $\sigma_{\phi z}$ is the cross term coupling vertical and axial motions. 

Integrating both sides of equation~(\ref{eqn:zjeans}), we derive our key equation for this paper: 
\begin{equation} 
\sigma_z^2(z) = \frac{1}{\nu(z)}\int_{0}^z \nu(z') \left[K_z(z') - \mathcal{T}(z') - \mathcal{A}(z') \right] dz' + \frac{C}{\nu(z)}
\label{eqn:kzlaw} 
\end{equation}  
where $C$ is a normalisation parameter. For $z=0$, we have that 
\begin{equation}
\sigma_z^2(0)\nu(0) = C,
\label{eqn:normC}
\end{equation}
and so $C$ simply sets the vertical velocity dispersion at $z = 0$. (Note that a similar but less general equation was derived recently in \citet{2012ApJ...746..181S}, equation (10).) We implement the constant $C$ as a parameter in our model that we will ultimately marginalise over. The alternative would be to calculate $C$ directly from equation~(\ref{eqn:kzlaw}), utilising the fact that as $z \rightarrow \infty$, $\nu(z) \rightarrow 0$, which gives us
\begin{equation} 
C =-\int_{0}^\infty \nu(z') \left[K_z(z') - \mathcal{T}(z') - \mathcal{A}(z') \right] dz'.
\label{eqn:mock_data_C}
\end{equation}
However this would mean that the derived value for $\sigma_z(z)$ at some finite $z$ would depend the properties of our mass model, tilt and axial terms not just in the range $[0, z]$ but also for the range $[0, \infty)$.   

Equation~(\ref{eqn:kzlaw}) is numerically appealing to solve since, unlike many previous methods \citep[e.g.][]{Garbari:2011dh,Garbari:2012ff}, it does not require any numerical differentiation and is therefore rather robust to noise in the data. Furthermore, equation~(\ref{eqn:kzlaw}) is valid for {\it any gravity theory} and can therefore be used as a constraint on alternative gravity models. In this sense, we follow the early pioneering work of \citet{1960BAN....15....1H} who attempted to also measure $K_z$ directly without reference to the Poisson equation.  

To connect the vertical acceleration $K_z$ to the surface mass density of the disc, however, we must specify a gravitational model. For standard Newtonian weak field gravity, this is given by the Poisson equation: 
\begin{equation}
\nabla^2 \Phi = \frac{\partial^2 \Phi}{\partial z^2} + \hspace{-3mm}\underbrace{\frac{1}{R}\frac{\partial V_c^2(R)}{\partial R}}_{\text{`rotation curve' term}:\,\,\mathcal{R}} \hspace{-3mm}= 4\pi G\rho
\label{eqn:poisson}
\end{equation}
where $V_c(R)$ is the circular speed (rotation) curve at radius $R$, and $\rho$ is the total matter density. Notice that for a flat rotation curve, $V_c = \mathrm{const.}$, and the `rotation curve' term $\mathcal{R}$ vanishes. If $\mathcal{R}$ does not vanish, then it appears as a shift (potentially as a function of height $z$) to the recovered $\rho_{\rm DM}$ that can be corrected for at the end of the calculation \citep{Garbari:2012ff}. Equation~(\ref{eqn:poisson}) can be rewritten as
\begin{equation}
\frac{\partial^2 \Phi}{\partial z^2} = 4 \pi G \rho(z)_{\rm eff}
\label{eqn:poisson_with_rhoeff}
\end{equation}
where 
\begin{equation}
\rho(z)_{\rm eff} = \rho(z) - \frac{1}{4 \pi G R} {\frac{\partial V_c^2(R)}{\partial R}}.
\label{eqn:rho_eff}
\end{equation}
The contribution of this term to the mass density profile can be quantified via the Oort constants $A$ and $B$ \citep{Binney2008}:
\begin{equation}
\frac{1}{4 \pi G R} \frac{\partial V_c^2(R)}{\partial R} = \frac{B^2 - A^2}{2 \pi G}.
\label{eqn:rot_term_oort_constants}
\end{equation}
Note that $|B|<|A|$, meaning that the terms in equation~(\ref{eqn:rot_term_oort_constants}) are negative. Thus the effective density $\rho(z)_{\rm eff}$ will be higher with the inclusion of a non-zero rotation curve term. If the rotation curve term is erroneously neglected it will be absorbed into the density profile, yielding an over-estimate of the DM density, assuming the baryonic contribution is well constrained. Taking the most precise values for $A$ and $B$ from Table~\ref{tab:oort_constants} \citep{2000AandA...354..522M} yields a value for equation~(\ref{eqn:rot_term_oort_constants}) of $\sim 0.1\, {\rm GeV}/{c^2}$, which is roughly a third of the expected local DM density \citep[e.g.][]{Read:2014qva}. For any accurate measure of the local DM density derived from real data we will need to incorporate this correction, but we leave this for future work. 

\begin{table}
\caption{Values of Oort constants. We include the F giants even though the errors for them are substantially larger to show that, within current uncertainties, the Oort constants $A$ and $B$ do not depend on stellar type.}
\begin{tabular}{l l c c}
\hline 
Source & Stellar& $A$ & $B$\\
&Type & \multicolumn{2}{c}{ [km\,s$^{-1}$kpc$^{-1}$] }\\
\hline
\cite{2000AandA...354..522M} & K-M giants & $14.5\pm 1.0$ & $-11.5 \pm 1.0$\\
\cite{2010MNRAS.409.1269B} & F giants & $14.85 \pm 7.47$ & $-10.85 \pm 6.83$\\
\cite{2011RMxAA..47..197B} & G giants & $14.05 \pm 3.28$ & $-9.30 \pm 2.87$\\
\hline
\end{tabular}
\label{tab:oort_constants}
\end{table}

Neglecting $\mathcal{R}$ and integrating both sides of equation~(\ref{eqn:poisson}), we derive the familiar result: 
\begin{equation}
\Sigma_z(z) = \frac{|K_z|}{2\pi G} 
\label{eqn:sigzkz}
\end{equation}
where $\Sigma_z(z)$ is the surface mass density of the disc.  

The overall flow of our method is now apparent. We model the tracer density $\nu(z)$, the mass density distibution $\rho(z) = \rho_{\rm DM}(z) + \rho_{\rm baryon}(z)$ (which gives us the $K_z$ term), the tilt term $\mathcal{T}(z)$, and the axial term $\mathcal{A}(z)$.  These elements each have a number of parameters, and so in total each model will have an $N$-dimensional parameter space. Specifying values for each of these parameters will then give us a quantitative values for each of these elements, which can then be inserted into equation~(\ref{eqn:kzlaw}) to derive $\sigma_z^2(z)$ for that parameter space point. We then compare $\nu(z)$ and $\sigma_z^2(z)$ (and $\sigma_{Rz}(z)$ as part of the tilt term model) to data via a $\chi^2$ test, and then change the values of our parameters.

Note that the only assumptions that goes into equation~(\ref{eqn:kzlaw}) are that the motions of stars obey the collionless Boltzmann equation, and that the galaxy is in dynamic equilibrium. The assumption of dynamic equilibrium can be negated somewhat by the presence of disequilibria (`wobbling') in the disc caused by e.g. the Sagittarius merger \citep{2011Natur.477..301P, 2013MNRAS.429..159G} or in part by the presence of spiral arms \citep{2014MNRAS.440.2564F}. However the impact of these effects on the measurement of the local DM density is estimated to be approximately 10\% \citep{2012ApJ...750L..41W, Read:2014qva}, less than the corrections arising from the tilt and rotation curve terms.

In practice further assumptions are necessary to model the individual components on the RHS of equation~(\ref{eqn:kzlaw}). However this method can accommodate almost any model for each of these elements -- the only strict requirement is that each element can be defined at the $z$-values corresponding to the centres of the bins used to calculate the tracer density and velocity dispersion. Hence we call this method `non-parametric' -- we can in principle have many more parameters for each model than there are data points. In the following subsections we describe how we model each element for this particular study. In Section \ref{subsec:Multinest} we discuss model selection using the Bayesian evidence, which could potentially allow us to compare alternate models.

As the galaxy is close to being axisymmetric in both the thin disc \citep{1998AJ....115.2384D, 2005ApJ...629..268H, 2009MNRAS.397.1286A, 2009ApJ...700.1794B, 2012A&A...547A..71P, 2013A&A...551A...9A} and the thick disc \citep{2012A&A...547A..70P, 2014ApJ...793...51S}, the axial term $\mathcal{A}(z)$ is expected to be small. For this study we assume complete axisymmetry, and thus take $\mathcal{A}(z) = 0$. If the data shows significant non-axisymmetry in the Milky Way, the axial term could be modeled in a similar way to the tilt term as described below. 

Note that if the axial term were non-negligible, in the context of the Milky Way, this would imply the presence of significant spiral arms or a bar. Such features are time dependent, implying that if $\mathcal{A}(z) \neq 0$, we should also worry about the time dependence of the gravitational potential $\Phi(t)$ and, therefore, also the partial time derivative of the distribution function $\partial{f}/\partial{t}$ \citep[e.g.][]{2016MNRAS.457.2569M}. It is not clear if there can exist a regime in which $\mathcal{A}(z) \neq 0$ and such time dependence can be ignored. Here, we simply note that since empirically $\mathcal{A}(z) \ll \mathcal{T}(z) \, \forall z$ \citep[e.g.][]{2012A&A...547A..70P, 2012A&A...547A..71P}, to leading order we can also safely neglect the time dependence of the distribution function, as we have already done. 

\subsection{Tracer density model}\label{subsec:tracer_model} 
To apply the Jeans equations we bin data to obtain the tracer density $\nu$ and the velocity dispersions $\sigma_{z}$ and $\sigma_{Rz}$. Thus at a bare minimum we only have to define $\nu$ and $\sigma_{Rz}$, and derive $\sigma_{z}$ via equation~(\ref{eqn:kzlaw}), at the bin centers, i.e. at a discrete set of $z$ values. 

For this work we model the tracer density as a sum of $N$ exponential discs:
\begin{equation}
\nu(z) = \sum_i \nu(0)_i \exp \left(-\frac{z}{h_i} \right),
\end{equation}
where for the $i^{\rm th}$ disc $\nu(0)_i$ is the tracer density at $z=0$ and $h_i$ is the scale height. The number of exponential disks can be increased until the Bayesian evidence shows no improvement (see Section \ref{subsec:Multinest}). This method avoids overfitting as each disc is smooth, while still giving freedom to describe more complex data. From this point onwards we consider only $z\geq 0$.  
 
\subsection{Baryon parameterisation}\label{subsec:Baryon_profile}
For this study we use a simple 2-parameter model to describe the baryon distribution\footnote{In the context of the values of the prior ranges given in Table~\ref{tab:prior_ranges} below the gravitational constant equals $G = 4.229\times10^{-6} {\rm km}^2\, {\rm kpc}\, \Msun^{-1} {\rm s}^{-2}$.}:
\begin{equation}
\rho_{\rm baryon}(z) = \frac{1}{4 \pi G} \left| \frac{K_{\rm bn}D_{\rm bn}^2}{(D_{\rm bn}^2 + z^2)^{1.5}} \right |.
\label{eqn:disc_density}
\end{equation} 
The $K_{\rm bn}$ parameter sets the mass of the disc, and has dimensions of acceleration, while $D_{\rm bn}$ controls the scale height of the disc, and has dimensions of length. Expressed in terms of the $K_z$ parameter of equation~(\ref{eqn:kzlaw}), the baryon profile becomes
\begin{equation}
K_{z,{\rm baryon}} = - \left [ \frac{K_{\rm bn} z}{\sqrt{z^2 + D_{\rm bn}^2}} \right ].
\label{eqn:Kz_baryon}
\end{equation}
While this model is not likely realistic for the Milky Way \citep{2006MNRAS.372.1149F, 2015ApJ...814...13M}, it has been applied to observational data by \cite{1989MNRAS.239..571K, 1989MNRAS.239..605K}, and also more recently by \cite{2013ApJ...772..108Z}. When applying our method to real data we will consider more sophisticated baryon models, but for this initial study where we are primarily interested in testing our methodology, equation \ref{eqn:Kz_baryon} will suffice. 

\subsection{Dark Matter parameterisation}\label{subsec:DM_profile}
The simplest way to parameterise the DM density profile $\rho_{\rm DM}$ is to assume it is constant with $z$, as done in previous work \citep[e.g.][]{1984ApJ...287..926B, 1989MNRAS.239..571K, 1989MNRAS.239..605K, 1998A&A...329..920C, Garbari:2011dh, Garbari:2012ff}. This assumption works well at low $z$: for a spherical NFW halo with a scale radius of 20 kpc the midplane value is correct within 10\% up to a height of $z\sim3$ kpc. For some analyses we also make this assumption, and set $\rho_{\rm DM}(z) = \rho_{\rm DM, const.}$.

However, this assumption does not allow for the exploration of several interesting effects such as a flattened, oblate halo or a dark disc. To accommodate such phenomena we add a dark disc (DD) on top of the constant DM. The dark disc is described using the same disc model as we use for the baryons, yielding a total DM profile:
\begin{equation}
\rho_{\rm DM}(z) = \rho_{\rm DM, const.} + \frac{1}{4 \pi G} \left| \frac{K_{\rm DD}D_{\rm DD}^2}{(D_{\rm DD}^2 + z^2)^{1.5}} \right |. 
\label{eqn:DM_profile}
\end{equation}
To ensure this disc does not simply become degenerate with the baryonic disc we give the scale height $D_{\rm DD}$ a higher prior range than that for $D_{\rm bn}$. In terms of the $K_z$ parameter of equation~(\ref{eqn:kzlaw}), the DM profile is
\begin{equation}
K_{z,{\rm DM}} = - \left [ 2Fz + \frac{K_{\rm DD} z}{\sqrt{z^2 + D_{\rm DD}^2}} \right ]
\label{eqn:Kz_DM}
\end{equation}
where $F = 2\pi G \rho_{\rm DM, const}$. This disc parameterisation of the non-constant DM profile can describe both the accreted dark disc and a flattened DM halo, and so henceforth we refer only to a `dark disc'.  

Additional dark disc terms could in principle be added to $\rho_{\rm DM}(z)$ with parameters $K_{{\rm DD}, n}$ and $D_{{\rm DD}, n}$, giving a total density profile of 
\begin{equation}
\rho_{\rm DM}(z) = \rho_{\rm DM, const.} + \frac{1}{4 \pi G} \sum_n \left| \frac{K_{{\rm DD},n}D_{{\rm DD},n}^2}{(D_{{\rm DD},n}^2 + z^2)^{1.5}} \right |.
\label{eqn:DM_profile_nDD}
\end{equation}
The prior ranges on the $n^{\rm th}$ dark disc parameters $K_{{\rm DD}, n}$ and $D_{{\rm DD}, n}$ can be set in relation to those of the $(n-1)^{\rm th}$ dark disc, e.g. requiring $K_{{\rm DD}, n} < K_{{\rm DD}, n-1}$, meaning each dark disc is less massive than the previous one. The number of dark discs to add could be determined from the data via the Bayesian evidence, with dark disc terms being added up until this degrades significantly. For this study however we limit ourselves to one dark disc term. 

\subsection{Tilt term} \label{subsec:Tilt_term}
\begin{figure}
\includegraphics[width=0.9\linewidth]{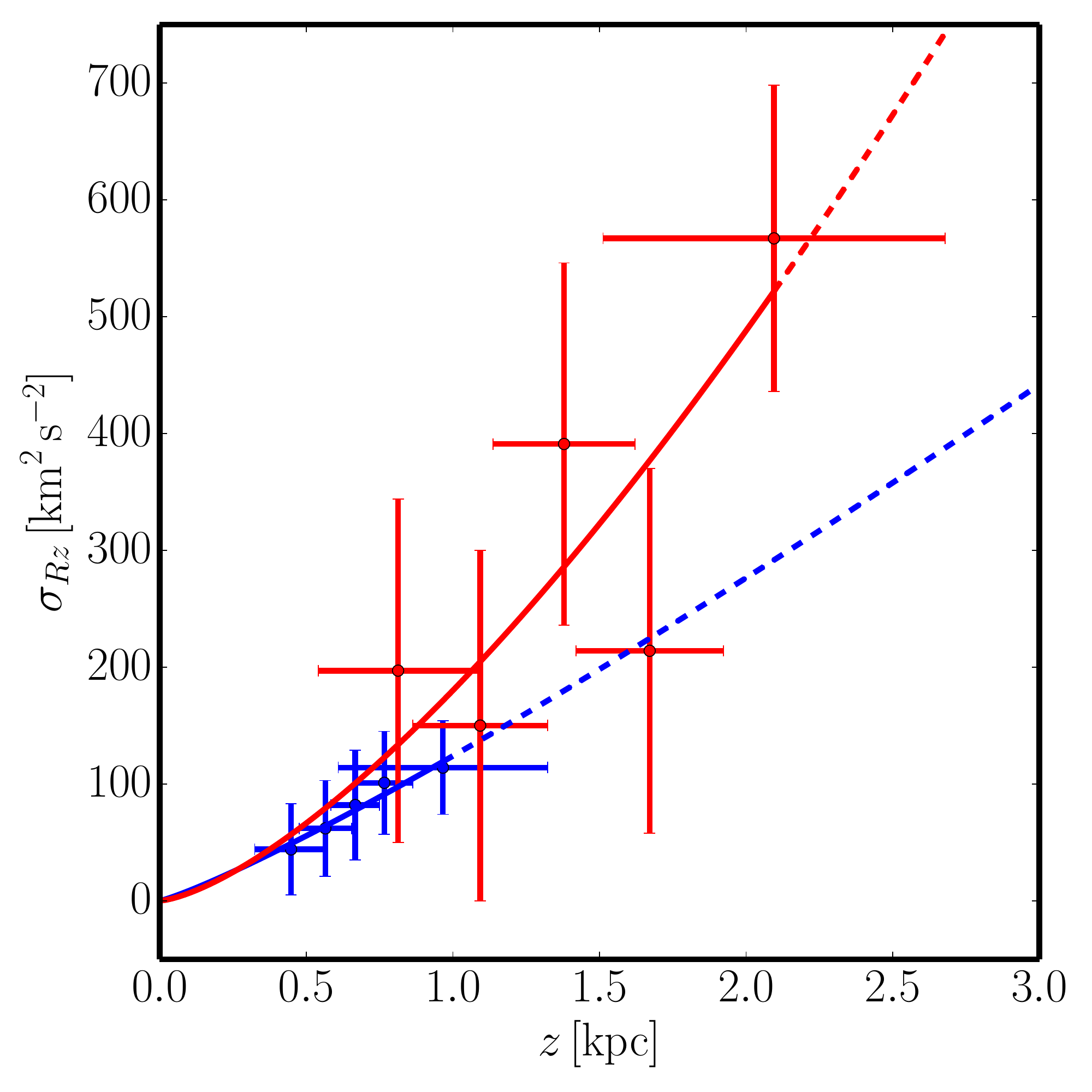}
\caption{$\sigma_{Rz}$ for two populations of G-type dwarf stars in the solar neighbourhood from the SDSS/\textit{SEGUE} survey \citep{2015MNRAS.452..956B}. The blue data points are from a younger, high metallicity population, with $\overline{[{\rm Fe/H}]} = -0.07$ and $\overline{[\alpha{\rm /H}]} = 0.11$, while the data points in red are from an older, low metallicity population with $\overline{[{\rm Fe/H}]} = -0.89$ and $\overline{[\alpha{\rm /H}]} = 0.34$. The blue and red lines are power laws in the form of equation~(\ref{eqn:sigmaRz}) fitted to the blue and red data points respectively. The older, lower metallicity stars probe further above the disc plane. Dashed lines indicate extrapolation from data.}
\label{fig:budenbender_sigRz}
\end{figure}

\begin{figure}
\includegraphics[width=0.9\linewidth]{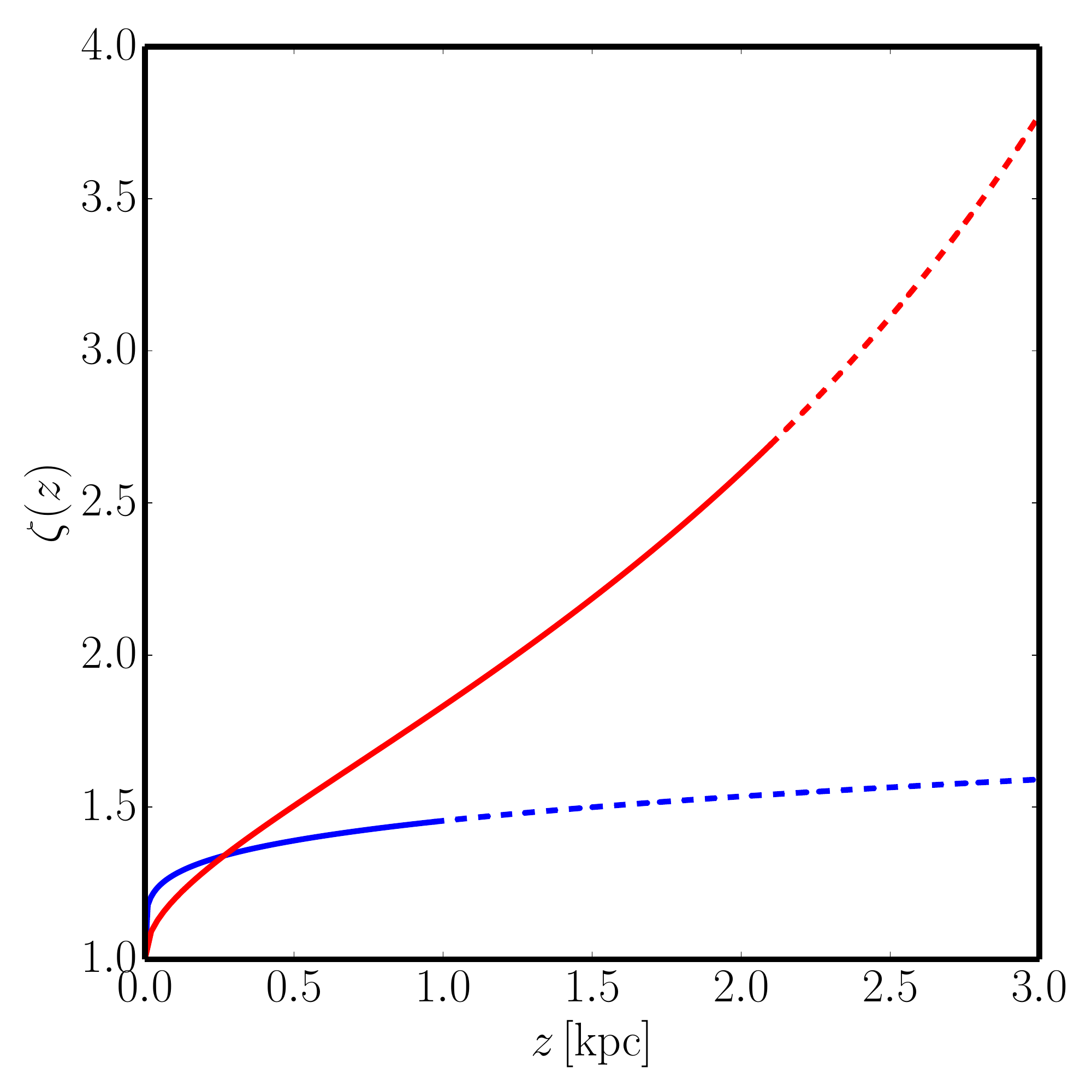}
\caption{The importance of tilt, as quantified by equation~(\ref{eqn:tilt_zeta}). The blue and red lines correspond to tilt terms derived from the high and low metalicity population fits of Fig.~\ref{fig:budenbender_sigRz}. The deviation arising from the tilt term increases more rapidly with height for the low metallicity population (red line), which probes the high-$z$ region most useful for determining the DM profile. Dashed lines indicate extrapolation from data. For this case $R_0 = R_1 = 2.5$ kpc., and $F = 267.65\, {\rm km}^2 \, {\rm kpc}^{-2}\, {\rm s}^{-2}$.}
\label{fig:budenbender_zeta_tilt}
\end{figure}

The tilt term from equation~(\ref{eqn:zjeans}) links radial and vertical motion. The importance of this term has been noted previously, e.g. \cite{1989MNRAS.239..571K, 2009ApJ...698.1110S, 2015MNRAS.452..956B}. Here we demonstrate that it is possible to deal with this term while remaining within our vertical, one-dimensional framework. Given the data quality currently available for the Milky Way we are required to make a well-motivated assumption about the radial form of $\mathcal{T}$, but in the Gaia-era we will be able to directly measure this from the data. 

We first assume that the radial profiles of the tracer density and the $(R,z)$-velocity dispersion are exponentials with scale radii of $R_0$ and $R_1$, respectively:
\begin{align}
\nu(R,z) &= \nu(z) \exp(-R/R_0),\\
\sigma_{Rz} (R,z) &= \sigma_{Rz}(z) \exp(-R/R_1).
\end{align}
With this assumption the tilt term becomes
\begin{equation}
\mathcal{T}(R,z) = \sigma_{Rz} (R,z) \left[\frac{1}{R} - \frac{1}{R_0} -\frac{1}{R_1} \right].
\label{eqn:tiltRzR0R1}
\end{equation}
If the disc were observed not to be an exponential then an alternative model could be easily applied at this stage. Indeed, similar but more complex models have featured previously in the literature, e.g. \citep{2014MNRAS.439.1231B}.

We then model the $(R,z)$-velocity dispersion as a power law at a given galactocentric radius $R$:
\begin{equation}
\sigma_{Rz}(R,z) = \left. A\left(\frac{z}{{\rm kpc}}\right)^n  \right |_R.
\label{eqn:sigmaRz}
\end{equation}
We take $R = R_\odot$, and also $R_0 = R_1$, simplifying the tilt term to a model decribed by the parameters $A$, $n$, and $R_0$:
\begin{equation}
\mathcal{T}(R_\odot, z) = \left. A \left(\frac{z}{{\rm kpc}}\right)^n \right |_{R_\odot} \left[\frac{1}{R_\odot} - \frac{2}{R_0} \right].
\label{eqn:tilt_term}
\end{equation}
Note that we are not affected by the assumption $R_0=R_1$ since these two parameters are trivially degenerate in equation \ref{eqn:tilt_term}. It remains the case that any observational constraints on $R_0$ and/or $R_1$ can be used as priors on equation \ref{eqn:tilt_term}, where these would constrain the term $2/R_0$. The description of $\sigma_{R,z}(z)$ of equation~(\ref{eqn:sigmaRz}) fulfils $\sigma_{R,z}(z=0) = 0$ by construction and fits remarkably well to different populations, as shown in Fig.~\ref{fig:budenbender_sigRz}. This figure presents $(z, \sigma_{Rz})$ data points for high and low metallicity populations (blue and red points respectively) from the SDSS/\textit{SEGUE} survey as analysed and presented in \cite{2015MNRAS.452..956B}, with a sign correction applied\footnote{Private communication with authors.}. The high metallicity sample has $\overline{[{\rm Fe/H}]} = -0.07$ and $\overline{[\alpha{\rm /H}]} = 0.11$, while the low metallicity sample has  $\overline{[{\rm Fe/H}]} = -0.89$ and $\overline{[\alpha{\rm /H}]} = 0.34$. Metallicity can be used as a proxy for age, with the high metallicity sample being younger than the older, low metallicity population \citep[e.g.][]{2011ApJ...737....8L}. 

Taking these data points we fit power laws models as per equation~(\ref{eqn:sigmaRz}), with parameters $A = 123.99$\,km$^2$ s$^{-2}$, $n =  1.16$ for the high metallicity population (blue) and $A =180.08$\,km$^2$ s$^{-2}$, $n=  1.44$ for the low metallicity population (red). The low metallicity population samples further above the disc plane and populates the canonical thick disc of the Milky Way. Populations such as this are more interesting for local DM searches as they allow us to probe higher $z$ regions where the baryon mass contribution begins to drop away leaving behind the DM halo. However as we go higher in the disc the tilt term becomes increasingly important, as illustrated by Figure \ref{fig:budenbender_zeta_tilt}, which shows the variable $\zeta(z)$:
\begin{equation}
\zeta(z) = \frac{K_{z,{\rm DM}}}{ K_{z,{\rm DM}} - \mathcal{T} }
\label{eqn:tilt_zeta}
\end{equation}
where $K_{z,{\rm DM}} = -2Fz$, the constant DM density term, with $F = 267.65\, {\rm km}^2 \, {\rm kpc}^{-2}\, {\rm s}^{-2}$ and $R_0 = R_1 = 2.5$ kpc, the same values as are used to later generate our mock data in Section \ref{sec:mock_data}. Compared to the thin disc population (blue line), the effects of the tilt term become important at much lower $z$ values. In short, to probe local DM we want to use thick disc stars probing higher-$z$ ranges, but the cost is that we must include the tilt term in our analyses. 

Ignoring the tilt term will always cause an underestimation of the local DM density, if all other components of the model such as baryons are held steady. Looking at equation~(\ref{eqn:tilt_term}), we note that $R_0 < R_\odot$ and $A >0$, meaning the tilt term $\mathcal{T}(R_\odot, z)$ is always negative. Then considering equation~(\ref{eqn:kzlaw}) we see that to fit to $\sigma_z^2(z)$ in the absence of the tilt term $\mathcal{T}(z')$, the $K_z(z')$ term, a negative term, is forced to become less negative in order to compensate. This requires a lower mass density, which if the baryon density profile is held constant, manifests itself as a decrease in the DM mass density.  

\subsection{Statistical analysis and \MultiNest}\label{subsec:Multinest}
\begin{table}
\caption{Prior ranges for parameters. Gaussian priors are described by a median $\mu$ and a standard deviation $\sd$.  Note that $\nu(0)$ and $\sigma_{z}(0)$ are the tracer density and velocity dispersion at $z=0$, while quantities subscripted with 0, such as $\nu_0$ and $\sd_{\nu, 0}$, are the values derived from data in the $0^{\rm th}$ bin, whose bin centre $z_0 > 0$. Tracer density $\nu(0)$ has units of $[{\rm stars}\, {\rm kpc}^{-3}]$. $K_{\rm bn}$ and $K_{\rm DD}$ terms have units $[{\rm km}^2 \, {\rm kpc}^{-1}\, {\rm s}^{-2}]$.}
\begin{tabular}{l r c c}
\hline
\multicolumn{2}{c}{Model Parameter} & Range {\it or} Gaussian $\mu$ \& SD & Type\\
\hline
Baryons & $K_{\rm bn}$ & $\mu=1500$, $\sd=150$ & Gaussian\\
& $D_{\rm bn}$ & $\mu=0.18$ kpc, $\sd=0.02$ kpc & Gaussian\\
\\
Constant DM & $\rho_{\rm DM}$ & $[2, 20] \times 10^{-3}\,\Msun \,{\rm pc}^{-3} $ & Linear\\
& & $[0.076, 0.796]\, {\rm GeV} \,{\rm cm}^{-3}$ & \\
\\
Dark Disc & $K_{\rm DD}$ & [0, 1500] & Linear\\
& $D_{\rm DD}$ & [1.5, 3.5] kpc & Linear\\ 
\\
Tilt term & $A$ & [0, 200] ${\rm km}^2\, {\rm s}^{-1}$ & Linear\\
& $n$ & [1.0, 1.9] & Linear\\
& $R_0$ & $\mu = 2.5$ kpc, $\sd=0.5$ kpc & Gaussian\\
\\
Tracer density & $\nu(0)$ & [0, $\nu_0$ + $2\times\sd_{\nu, 0}$] & Linear\\
& $h$ & [0.4, 1.4] kpc & Linear\\
\\
Velocity disp. & $\sigma_{z}(0)$ & [$\sigma_{z,0} - 2 \times \sd _{\sigma_z, 0}$,  & Linear \\  
& & $\sigma_{z,0} + 2\times \sd_{\sigma_z, 0}$] km s$^{-1}$ & \\
\hline
\end{tabular}

\label{tab:prior_ranges}
\end{table}
The model outlined above gives us an $N$-dimensional parameter space. To explore this parameter space and derive limits on the various observables we adopt nested sampling \citep{skilling2006} as implemented in the publicly available \MultiNest code \citep{Feroz:2007kg, Feroz:2008xx, 2013arXiv1306.2144F}. \MultiNest is a tool for Bayesian inference and parameter estimation. Bayes Theorem is
\begin{equation}
P(\theta | D, \mathcal{M}) = \frac{P(D | \theta, \mathcal{M}) P(\theta | \mathcal{M})}{P(D|\mathcal{M})}
\label{eqn:BayesTheorem}
\end{equation}
where $\mathcal{M}$ is the given model, $\theta$ is the set of parameters for that model, and $D$ is the data. The left hand side, $P(\theta | D, \mathcal{M})$ is known as the \textit{posterior}, while the three terms on the right are the \textit{likelihood} $P(D | \theta, \mathcal{M}) = \mathcal{L}(\theta)$, the \textit{prior} $P(\theta | \mathcal{M})$, and the \textit{Bayesian evidence} $P(D|\mathcal{M})$.

The Bayesian evidence, a.k.a. the marginal or model likelihood, is a measure of how well the model performs given the data, and can be expressed as
\begin{equation}
\mathcal{Z} = P(D|\mathcal{M}) = \int P(D|\theta,\mathcal{M})P(\theta|\mathcal{M}) d\theta.
\label{eqn:bayesian_evidence}
\end{equation}
The performance of two different models given the same data can be compared using the \textit{Bayes factor}:
\begin{equation}
B_{01} = \frac{P(D|\mathcal{M}_0)}{P(D|\mathcal{M}_1)}.
\label{eqn:bayes_factor}
\end{equation}

Assuming Gaussian errors it is possible to derive an empirical scale relating the Bayes factor to the strength of evidence for one model over another, as done in \citep{Trotta:2008qt}. There, a $|\ln B_{01}|$ value of less than 1 is considered inconclusive, while values of 1.0, 2.5 and 5.0 are considered to give weak, moderate and strong evidence, respectively.
 
\MultiNest takes a given prior probability distribution and likelihood function and calculates the posterior distribution and Bayesian evidence. Our likelihood function is based on the $\chi^2$ distribution:
\begin{equation}
\mathcal{L}(\theta) = \exp{\left( \frac{-\chi^2}{2} \right)}.
\label{eqn:chi2_likelihood}
\end{equation}
The value of $\chi^2$ is simply
\begin{equation}
\chi^2 = \chi^2_\nu + \chi^2_{\sigma^2_{z}} + \chi^2_{\sigma_{Rz}}, 
\label{eqn:chi2_sum}
\end{equation}
where
\begin{align}
\chi^2_\nu &= \sum_j \frac{(\nu_{{\rm data},j} - \nu_{{\rm model},j})^2}{\sd_{\nu,j}^2}, \label{eqn:chi2_nu}\\
\chi^2_{\sigma_z} &= \sum_j \frac{(\sigma_{z,{\rm data},j} -\sigma_{z,{\rm model},j} )^2}{\sd_{{\sigma^2_z},j}^2}, \label{eqn:chi2_sigz}\\
\chi^2_{\sigma_{Rz}} &=\sum_j \frac{(\sigma_{Rz,{\rm data},j} -\sigma_{Rz,{\rm model},j} )^2}{\sd_{{\sigma_{Rz}},j}^2}. \label{eqn:chi2_sigRz}
\end{align}
Note that if the reconstruction model does not contain a tilt term (\textit{e.g.} $\mathcal{T} = 0$) then $\chi^2_{\sigma_{Rz}}$ is set to zero.

Table \ref{tab:prior_ranges} shows the prior ranges used for our analyses.  
We derive credible regions (CRs) for the DM density parameters by taking its posterior distribution and marginalising over the remaining parameters. As outlined above our model has several components that can be turned on or off, such as the dark disc. Using the Bayesian evidence as calculated by \MultiNest it is potentially possible to perform model comparison to determine which reconstruction model best fits the data. This idea will be explored in greater depth in subsequent studies.

\section{Mock data sets}\label{sec:mock_data}
In this paper we apply our method only to mock data in order to hone and verify it. This mock data is `as good as it gets', in the sense that it has no measurement errors, nor observational biases added to it, and is drawn from relatively simple, known distribution functions. While dynamically realistic `N-body' mocks are preferred, it has already been shown that 1D Jeans analyses are robust to the presence of local non-axisymmetric structure in the disc \citep{Garbari:2011dh}. Furthermore, N-body mocks are expensive; even state-of-the-art simulations do not approach the local sampling expected from Gaia. Finally, the expense of building well sampled N-body mocks makes it challenging to explore a large parameter space of models, including models with and without tilt, or with and without a dark disc. For these reasons, we focus here on simpler mock data, similarly to the Read (2014) review. We will consider more dynamically realistic mocks, and mocks that give a faithful representation of the expected Gaia data uncertainties in future work. 

We generate here a variety of mock data sets as outlined in Table~\ref{tab:mockdataparams}, with different samplings ($10^4$, $10^5$, and $10^6$ tracer stars), with and without a tilt term, and also with either no dark disc, a dark disc ({\tt dd}), or a more massive `big dark disc' ({\tt bdd}). We assume the axial term $\mathcal{A}$ and the rotation curve term $\mathcal{R}$ are zero. Our simple baryonic disc model is set up to mimic the real Milky Way, with a scale height parameter of $D_{\rm bn} = 0.18$\,kpc and a surface density of $\Sigma_{\rm bn} = 55.53\, \Msun \, {\rm pc}^{-2}$, similar to those measured near the solar position \citep{2006MNRAS.372.1149F, Read:2014qva, 2015ApJ...814...13M}. The value of the $F$ parameter (see equation \ref{eqn:Kz_DM}) corresponds to a DM density of $\rho_{\rm DM, const} = 10 \times 10^{-3}\, \Msun \,{\rm pc}^{-3} = 0.38\, {\rm GeV} \,{\rm cm}^{-3}$. For each scenario we generate 20 mock data sets, allowing us to explore the effects of poisson noise over a range of realisations.

The mock data consist of a list of stars each with three pieces of data: the position $z$, the vertical velocity $v_z$, and the product of the vertical and radial velocities, $v_R v_Z$. The first element, $z$, is generated by drawing randomly from an exponential tracer distribution with scale height $h$:
\begin{equation}
\nu(z) = \exp \left(-\frac{z}{h} \right).
\label{eqn:exp_nu_for_mocks}
\end{equation}
This gives us our list of stellar positions. 

To derive the velocites for the mock catalogue we first must define a mass model and tilt. Taking the same parameterisations as described in Sections \ref{subsec:Baryon_profile}, \ref{subsec:DM_profile}, and \ref{subsec:Tilt_term} for baryons, DM, and tilt, respectively, we set their parameters as per Table~\ref{tab:mockdataparams}. This allows us to calculate $K_z$ and $\mathcal{T}(z)$. Using equation~(\ref{eqn:mock_data_C}) and its associated assumptions we can calculate $C$. We then use equation~(\ref{eqn:kzlaw}) to derive $\sigma_z(z)$. For each star we take its position $z'$, find the value of $\sigma_z(z')$, and then draw a velocity from a Gaussian centred on $v_z = 0$ and with variance $\sigma_z^2(z')$. 

To generate $v_R v_Z$ mock data, when necessary, we take the $A$, $n$, and $R_0$ parameters listed in Table~\ref{tab:mockdataparams} and calculate the $\sigma_{Rz}(z)$ profile via equation~(\ref{eqn:sigmaRz}). Taking each star's position $z'$, we calculate $\sigma_{Rz}(z')$, and draw a value of $v_R v_z$ from a Gaussian centered on $v_R v_z = 0$ and with variance of $\sigma_{Rz}^2(z')$.

\begin{table}
\caption{Mock data parameters. {\tt \_X} is the number of stars sampled, \textit{e.g.} $10^4$, $10^5$, $10^6$, and {\tt \_M} is the mock number, ranging from 0 to 19. Empty spaces indicate that a certain mock data set does not include that element. The baryon model corresponds to a baryonic surface density of $\Sigma_{\rm bn} = 55.53\, \Msun \, {\rm pc}^{-2}$, while the $F$ parameter corresponds to a constant DM density of $\rho_{\rm DM, const} = 10 \times 10^{-3}\, \Msun \,{\rm pc}^{-3} = 0.38\, {\rm GeV} \,{\rm cm}^{-3}$. $K_{\rm bn}$ and $K_{\rm DD}$ terms have units $[{\rm km}^2 \, {\rm kpc}^{-1}\, {\rm s}^{-2}]$, while $F$ has units $[{\rm km}^2 \, {\rm kpc}^{-2}\, {\rm s}^{-2}]$.}
\begin{tabular}{l r c c c c c}
\\
\\
\\
\\
\\
\\
\\
\\
&& \begin{rotate}{90} {\tt thick\_X\_M} \end{rotate} 
&\begin{rotate}{90}{\tt thick\_dd\_X\_M} \end{rotate}
&\begin{rotate}{90}{\tt thick\_bdd\_X\_M} \end{rotate}
&\begin{rotate}{90}{\tt thick\_tilt\_X\_M} \end{rotate}
&\begin{rotate}{90}{\tt thick\_bdd\_tilt\_X\_M}\end{rotate}\\
\hline
Tracer & $h$ [kpc] & \multicolumn{5}{c}{\ruleline{0.9}} \\
density & & & & & &\\
\hline
Potential & $K_{\rm bn}$ & \multicolumn{5}{c}{\ruleline{1500}}\\ 
& $D_{\rm bn}$ [kpc] & \multicolumn{5}{c}{{\ruleline{0.18}}}\\
& $F$ & \multicolumn{5}{c}{{\ruleline{267.65}}}\\
\\
\hline
Dark Disc & $K_{\rm DM}$ & & 300 & 900 & & 900\\
& $D_{\rm DD}$ [kpc]  & & 2.5 & 2.5 & & 2.5 \\
\\
\hline
Tilt Term & $A$ [km$^2$ s$^{-2}$] & \, &  & &  180.08 & 180.08\\
& $n$ & & & & 1.44 & 1.44\\
& $R_0$ [kpc] & & & & 2.5 & 2.5 \\
\\
\hline
\end{tabular}
\label{tab:mockdataparams}
\end{table}

\section{Results}\label{sec:Results}
Here we present the results of our scans. We first investigate how the precision of the reconstruction varies with different numbers of tracer stars. We then look at the effects of the tilt term and the dark disc. For certain parts of these analyses the model used to generate the mock data and the model used to reconstruct the mock data are not the same -- this is done to investigate the effects of ignoring terms (as in the case of the tilt term) or using incorrect models. 

In the following figures we plot the marginalised posterior for the DM density $\rho_{\rm DM} (z)$, showing 68\%, 95\%, and 99.7\% CRs in dark, medium and light shading, respectively. The mock data profile is shown by the solid black line, while the median of the posterior distribution is shown by a solid line in the same colour as the CR. Binning of stars to calculate $\nu(z)$, $\sigma_z(z)$, and $\sigma_{Rz}(z)$ is performed such that each bin contains an equal number of stars. For this study we use 20 bins, and in Appendix \ref{app:binning} we briefly explore the effects of changing the number of bins used. For plots with constant DM density in both mock data set and reconstruction model we plot all 20 mock data sets together, while for non-constant DM density in either mock or reconstruction we show one representative figure in this section, and show the full set of figures in Appendix \ref{app:dd_grids} (Figs. \ref{fig:grid_MockNoDD_ModelDD} to \ref{fig:grid_MockBDD_MockTilt_ModelTilt_ModelDD}).

Our method and code is set up to describe the tracer density as a sum of exponentials. To determine the necessary number of exponentials required to describe the data we can use the Bayes factor as described earlier in Section \ref{subsec:Multinest}. Test reconstructions give a Bayes factor of 1.5 when comparing one exponential to two, and 3.3 when comparing one exponential to three, with one exponential favoured in both cases. This is to be expected given that the mock data is generated using a single exponential. Given this result we use a single exponential for all subsequent reconstructions. 

\subsection{Sampling}\label{subsec:results_sampling}
\begin{figure*}
\includegraphics[width=\textwidth,clip]{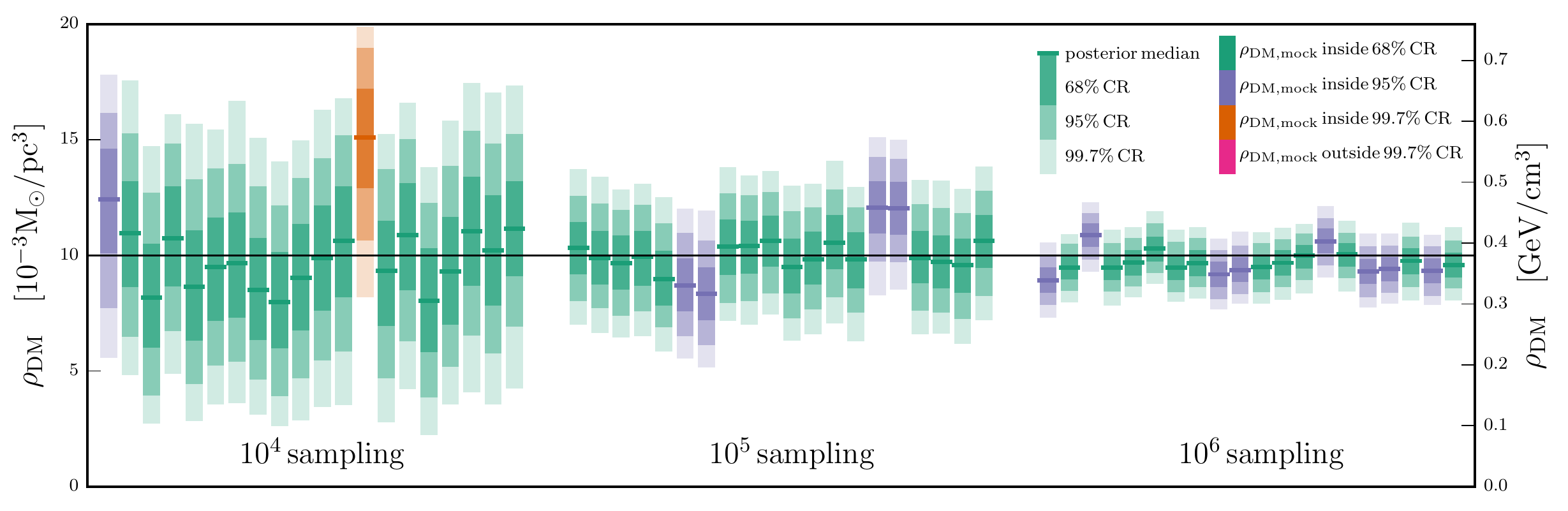}
\caption{Determination of the DM density profile for varying numbers of tracer stars. These plots show marginalised posteriors for $\rho_{\rm DM}(z) = \rho_{\rm DM, const}$ for the 20 mock data sets generated for each sampling level. Dark, medium, and light shading show the 68\%, 95\%, and 99.7\% credible regions (CRs), respectively. Green, purple, and orange colouring indicates that the 68\%, 95\%, or 99.7\% CR respectively contains the correct answer. The median value of each posterior is shown by a solid line in green, purple, or orange, while the DM density value used to generate the mock data is shown as a solid black line across the entire plot. The mock data and reconstruction models used contain no tilt term or dark disc terms. As the number of tracer stars is increased from $10^4$ to $10^6$ the credible regions for $\rho_{\rm DM, const}$ naturally shrink around the mock data value.}
\label{fig:sampling}
\end{figure*}

Fig.~\ref{fig:sampling} shows reconstructions of the local DM density for using varying numbers of tracer stars. The mock data sets used here are the most simple case, {\tt thick\_X\_M} as described in Table~\ref{tab:mockdataparams}, and has no dark disc or tilt added. As expected the credible regions shrink as the number of tracer stars is increased from $10^4$ stars up to $10^6$ stars. The SDSS survey has sampling on the order of $10^4$ stars \citep{2013ApJ...772..108Z, 2015MNRAS.452..956B} while data from Gaia will give upwards of $10^6$ tracer stars.

\subsection{Tilt}\label{subsec:results_tilt}
\begin{figure*}
\includegraphics[width=\textwidth,clip]{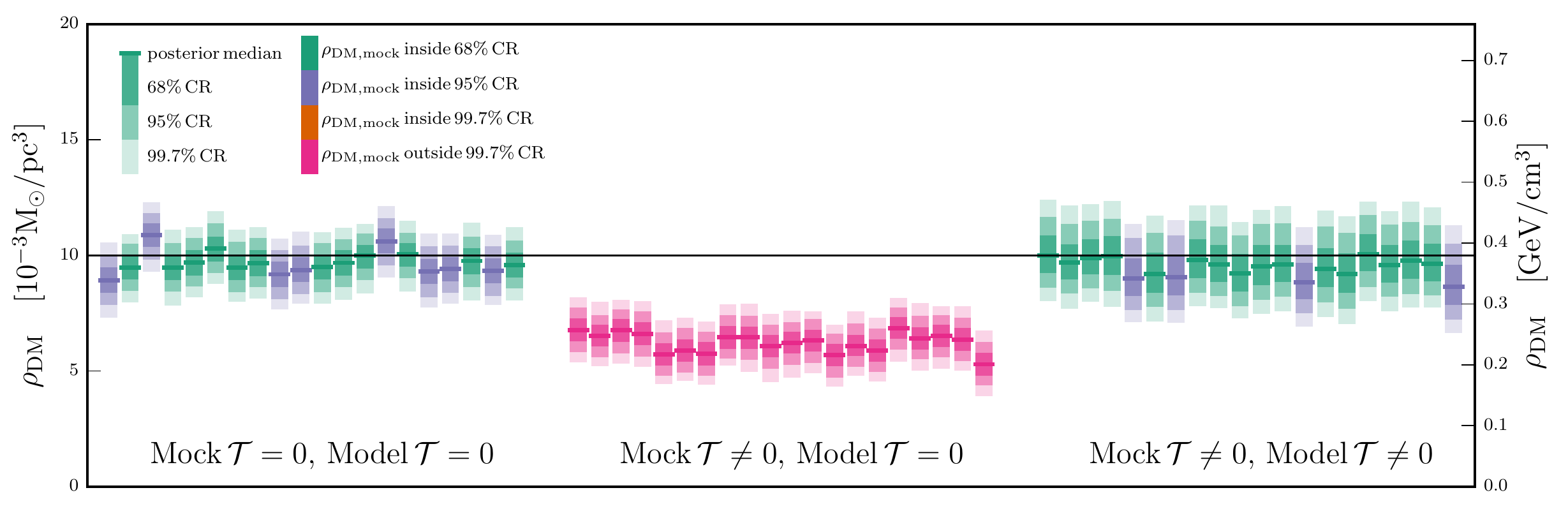} 
\caption{Determination of the DM density profile using $10^6$ tracer stars and exploring the effects of including or neglecting the tilt term in the mock data sets and reconstruction models. These plots show marginalised posteriors for $\rho_{\rm DM}(z) = \rho_{\rm DM, const}$ for the 20 mock data sets generated for each tilt scenario. Dark, medium, and light shading show the 68\%, 95\%, and 99.7\% credible regions (CRs) respectively. Green, purple, and orange colouring indicates that the 68\%, 95\%, or 99.7\% CR respectively contains the correct answer, while pink colouring indicated that the correct answer lies outside even the 99.7\% CR. The median value of each posterior is shown by a solid line in green, purple, or orange, while the DM density value used to generate the mock data is shown as a solid black line across the entire plot. The left set of mocks shows the same result as the right set of Fig.~\ref{fig:sampling}, with no tilt term in mock data or reconstruction model. The centre set has a tilt term in the mock data but not in the reconstruction model, yielding a systemic underestimation of $\rho_{\rm DM}$. The right set has a tilt term in both mock data and reconstruction model, demonstrating that our method can successfully deal with the tilt term.} 
\label{fig:tilt_1E6}
\end{figure*}

In Fig.~\ref{fig:tilt_1E6} we explore the effects of the tilt term. The left hand set of CRs in Fig.~\ref{fig:tilt_1E6} are the same as the right hand set of CRs from Fig.~\ref{fig:sampling}, with no tilt term in the mock data or reconstruction. In the centre set of Fig.~\ref{fig:tilt_1E6} however, the mock data {\tt thick\_tilt\_1e6\_M} has the tilt term turned on, but the analyses are performed with the tilt term set to zero. This illustrates the danger of ignoring tilt as discussed earlier in Section \ref{subsec:Tilt_term}: the reconstructions return narrow credible regions, but as expected they systematically underestimate the value of $\rho_{\rm DM}$, with the true $\rho_{\rm DM}$ lying outside even the 99.7\% CRs. This underestimation however is remedied by turning on the tilt term in the model, as shown in the right hand set CRs in Fig.~\ref{fig:tilt_1E6}, where the correct DM density always at least within the 95\% credible region. The inclusion of extra parameters to describe the tilt term does however increase the size of the CRs.

\begin{figure}
\includegraphics[width=\linewidth,clip]{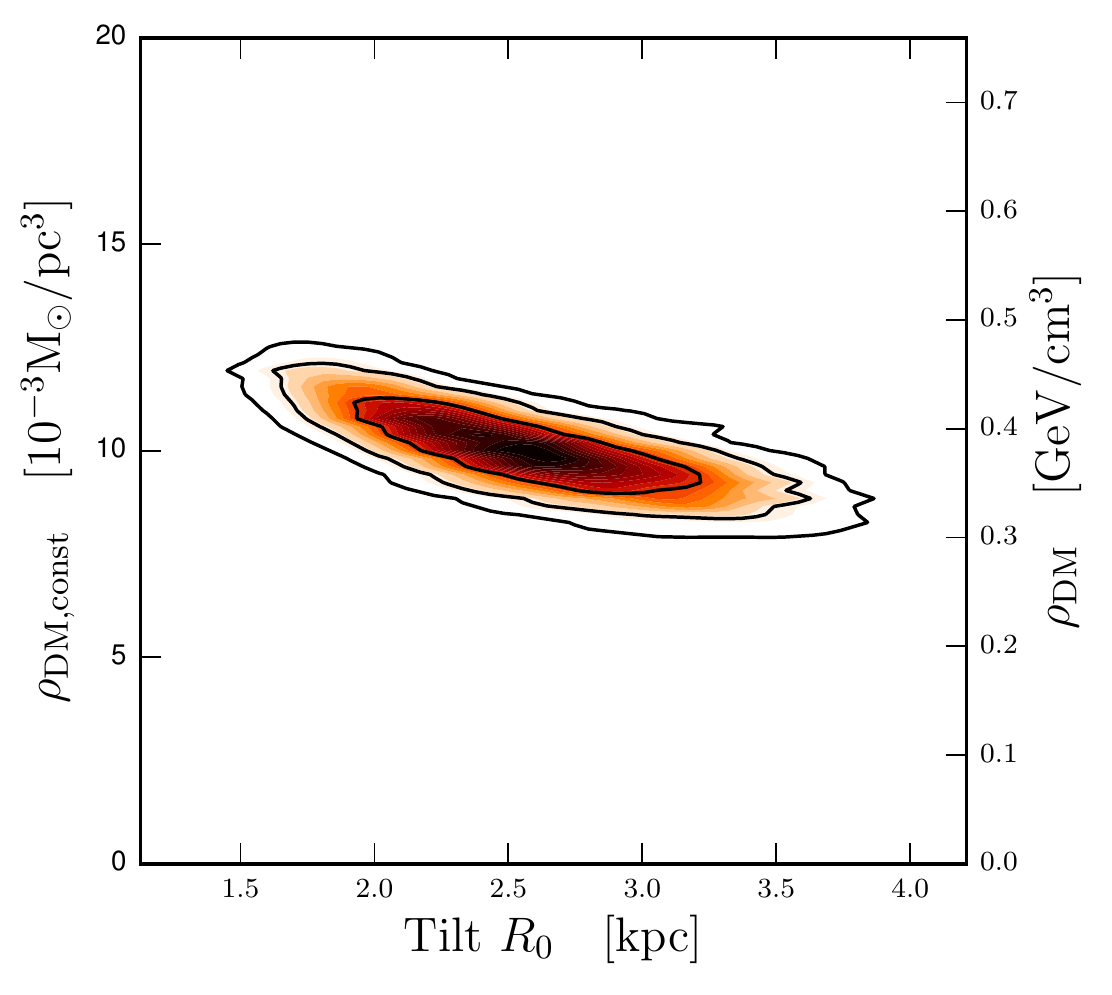}
\caption{2D marginalized posterior of the constant DM density $\rho_{\rm DM}(z) = \rho_{\rm DM, const}$ and the $R_0$ parameter of the tilt term (equation \ref{eqn:tiltRzR0R1}), generated from mock data set {\tt thick\_tilt\_1e6\_0} reconstructed using a model with tilt. Contours show the 68\%, 95\%, and 99.7\% CRs. Marginalization and plotting performed using {\sc Barrett} (Sebastian Liem, private communication).}
\label{fig:post_rhoDM_R0}
\end{figure}

Just as our determination of $\rho_{\rm DM}(z)$ is dependent on the tilt term, the tilt term is in turn dependent on its input parameters, $A$, $n$, and $R_0=R_1$. While we have been able to fit for $A$ and $n$ using G-type dwarf data (section \ref{subsec:Tilt_term}), for $R_0$ we have taken the canonical value of $R_0 = 2.5 \pm 0.5$ kpc from \citep{Binney2008}, and further made the assumption that $\sigma_{Rz}(R,z)$ has the same radial profile as $\nu(R,z)$ (i.e. $R_0 = R_1$). When using only a single population, determination of $R_0$ and $R_1$ will be important, as illustrated in Figure \ref{fig:post_rhoDM_R0}. This figure shows the 2D marginalized posterior for $\rho_{\rm DM, const}$ and $R_0$ for a reconstruction of mock data set {\tt thick\_tilt\_1e6\_0} with a model including tilt, and demonstrates the degeneracy that exists between $\rho_{\rm DM, const}$ and $R_0 = R_1$. If using multiple tracer population, each will have different $R_0$ and/or $R_1$, but will all have their motions dictated by the same potential. This will help us break the degeneracy between $R_0$, $R_1$, and $\rho_{\rm DM}$. Further, with the advent of Gaia data we will be able to directly measure the radial profile of $\sigma_{Rz}$ and $\nu(R,z)$ for a given set of tracer stars.

\subsection{Dark Disc}\label{subsec:results_darkdisc}
\begin{figure*}
\includegraphics[width=\linewidth, clip]{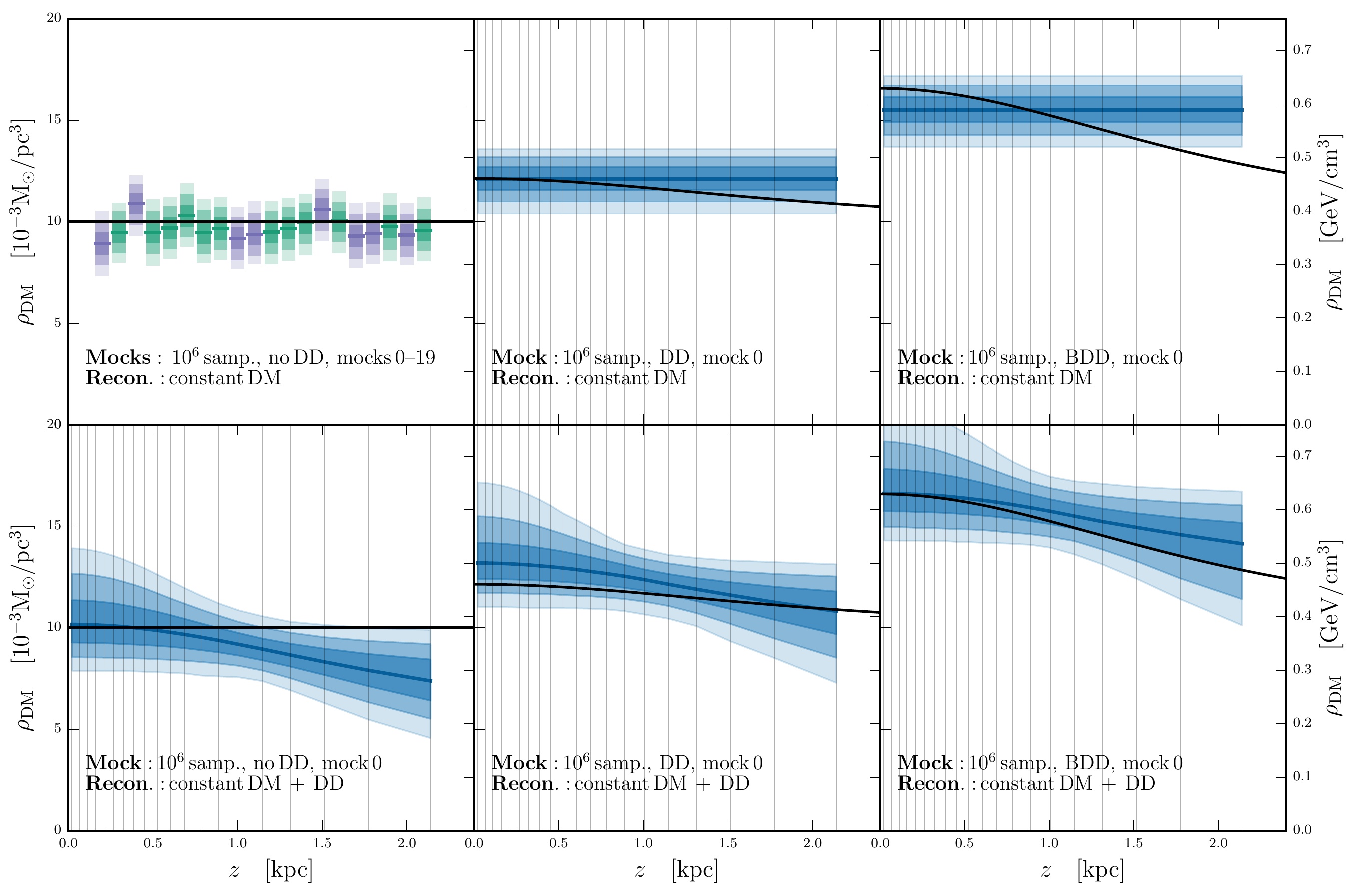} 
\caption{Determination of the DM density profile using $10^6$ tracer stars and exploring the effects of including or neglecting a dark disc in the mock data sets and reconstruction models. For comparison the top left panel shows the same reconstructions as the right hand set from Fig.~\ref{fig:sampling}, i.e. basic mock data sets with no DD, reconstructed with a constant DM density. The remaining panels each show one representative mock and reconstruction, with the full set of mocks and reconstructions given in Appendix \ref{app:dd_grids}.  These panels show marginalised posteriors for $\rho_{\rm DM}(z)$, with dark, medium, and light shading indicating the 68\%, 95\%, and 99.7\% CRs respectively. The median value of the posterior is shown by the solid blue line, while the DM density profile used to generate the mock data is shown as a solid black line. From left to right the columns show reconstructions of mock data sets containing no dark disc ($\rho_{\rm DM, const}$ only), a dark disc (DD), or a `big' dark disc (BDD). The top centre and top right panels show the determination of a constant DM density from the DD and BDD mock data sets respectively. The bottom row shows the reconstruction of each of the mock data sets using a model with a constant DM term and a dark disc term. Light grey vertical lines indicate the bin centres.}
\label{fig:darkdisc1E6}
\end{figure*}

Fig.~\ref{fig:darkdisc1E6} shows mock data set {\tt thick\_1e6\_0-19}, {\tt thick\_dd\_1e6\_0}, and {\tt thick\_bdd\_1e6\_0} reconstructed using models with and without a dark disc component. The top left panel shows the same CRs as seen in the left hand set o Fig.~\ref{fig:tilt_1E6} and the right hand set of Fig.~\ref{fig:sampling}.   

The left column of Fig.~\ref{fig:darkdisc1E6} shows the reconstruction of mock data sets with a constant DM density profile; mocks 0-19 in top left and mock 0 in bottom left. The reconstruction in the top panel uses a model with a constant DM density, while the bottom panel uses a model with an additional dark disc component. The dark disc reconstruction exhibits a disc structure even though the correct answer is constant $\rho_{\rm DM}$. This is likely due to a bias in the hyper-volume set by the priors -- the prior range on the dark disc parameters goes between no dark disc ($K_{\rm DD} = 0$) and a maximal dark disc ($K_{\rm DD} = 1500$), and thus the bulk of the parameter space features at least some dark disc. There is no `negative dark disc' to counteract this effect and push the mean of the prior range back to no dark disc.

In the centre and right columns of Fig.~\ref{fig:darkdisc1E6} we reconstruct mock data sets with a dark discs of different masses: {\tt thick\_dd\_1e6\_0} and {\tt thick\_bdd\_1e6\_0} (the `big dark disc'). A constant DM density reconstruction (top row) is able to contain the {\tt thick\_dd\_1e6} dark disc within the 95\% credible region almost to the last bin, but fails to contain the big dark disc beyond $z=1.3$ kpc. Adding a dark disc term allows the reconstruction to fit to the mock data DM profile nicely, as shown in the bottom centre and bottom right panels of Fig.~\ref{fig:darkdisc1E6}. 

When working with real data however, we will not have the luxury of knowing the DM profile underlying the data. The black mock data model line will not be there. In subsequent studies we will explore the potential use of the Bayesian evidence, as calculated by \MultiNest, to determine if a dark disc is justified by observational data. 

\subsection{Tilt and Dark Disc} \label{subsec:tilt_dd}
\begin{figure}
\includegraphics[width=\linewidth,clip]{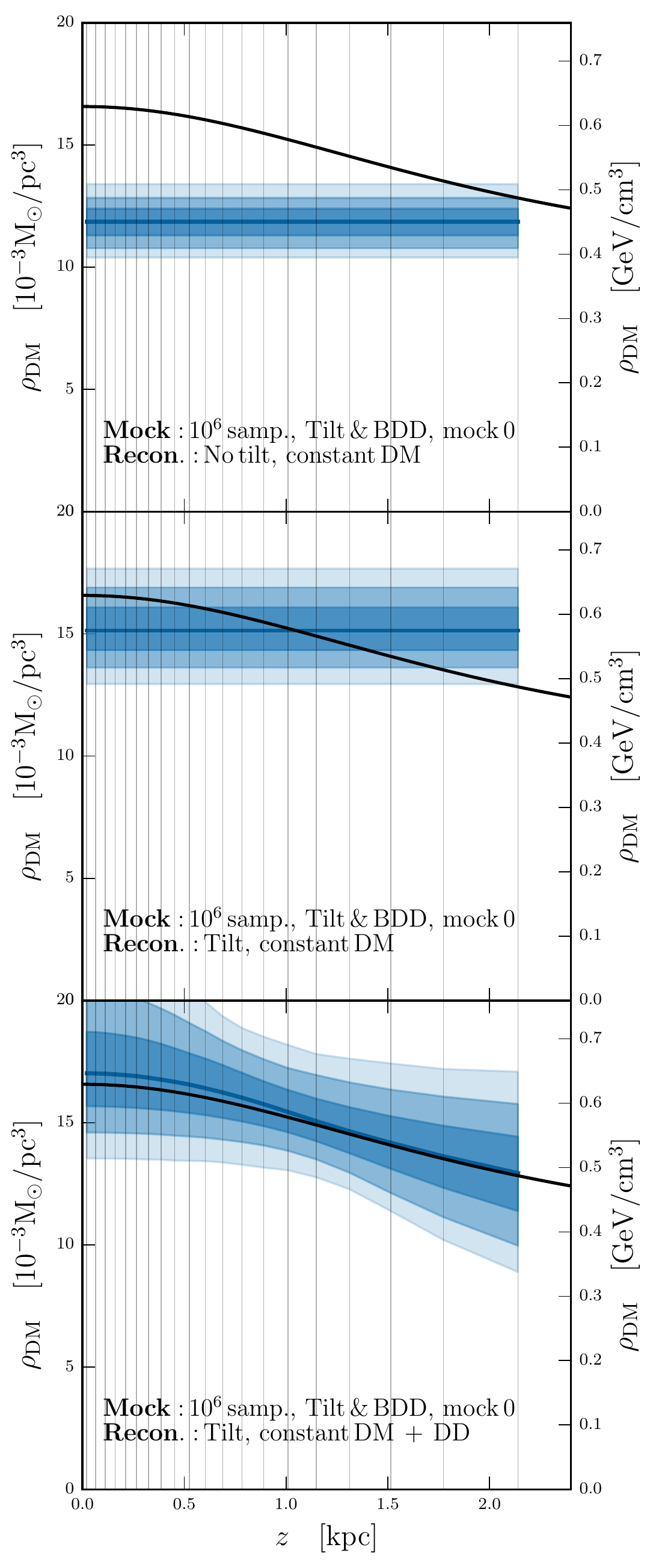}\\
\caption{Determination of the DM density profile using $10^6$ tracer stars with a combination of a tilt term and a dark disc in the mock data ({\tt thick\_bdd\_tilt\_1e6\_0}) and using a variety of reconstruction models. These plots show marginalised posteriors for $\rho_{\rm DM}(z)$, with dark, medium, and light shading indicating the 68\%, 95\%, and 99.7\% CRs respectively. The median value of the posterior is shown as the solid blue line, while the DM density profile used to generate the mock data is shown as a solid black line.}
\label{fig:DD_tilt_1E6}
\end{figure}

Here we combine the two elements discussed in previous sections, the tilt term and the dark disc. Fig.~\ref{fig:DD_tilt_1E6} shows reconstructions of the mock data set {\tt thick\_bdd\_tilt\_1e6\_0}. In the top panel the reconstruction model contains neither dark disc nor tilt term. Again we see the same effects as we did previously. The missing tilt term yields a consistent underestimation of the DM density, and the constant DM density envelope is too narrow to encompass the density range of the dark disc. The consistent underestimation can remedied by adding a tilt term to the reconstruction model, as shown in the middle panel, where the reconstruction model has a tilt term included. Both problems can be resolved in tandem by using a reconstruction model with both tilt and dark disc terms, as shown in the bottom panel of Fig.~\ref{fig:DD_tilt_1E6}.

\section{Conclusions} \label{sec:conclusions}
In this paper we have presented a new method of determining the vertical DM density profile, and thus the local DM density. The key equation of this method, equation~(\ref{eqn:kzlaw}), depends only on the assumption of dynamical equilibrium. In practice further assumptions are made to describe the components going in to equation~(\ref{eqn:kzlaw}), however the scope of possible models is wide, and in principle model selection using the Bayesian evidence can be used to determine the best model. We leave exploration of this last aspect to future work.  

The importance of the tilt term has been previously noted \cite{1989MNRAS.239..571K, 2009ApJ...698.1110S, 2015MNRAS.452..956B}, and to derive an accurate value for the local DM density it is vital that our method be able to deal with this term. The baryonic contribution to the galactic density profile drops rapidly as $z$ increases, leaving DM as the dominant component. Thus to probe the DM density profile sampling thick disc stars that travel higher above the plane is preferable. For these populations the tilt term is more important, as illustrated in Fig.~\ref{fig:budenbender_zeta_tilt}. Failure to include the tilt term in the analysis leads to a systematic underestimation of the local DM density, as explained in Section \ref{subsec:Tilt_term} and demonstrated in Section \ref{subsec:results_tilt}. 

One of the novel aspects of our method is that it can deal with the tilt term while remaining within the confines of the one-dimensional $z$-direction Jeans equation, which can be seen in Fig.~\ref{fig:tilt_1E6}. With only the data currently available for the Milky Way, this requires several well motivated assumptions, as described in Section \ref{subsec:Tilt_term}. However, with data from Gaia we will be able to directly measure the radial profile of the tilt and tracer density, removing the need for such assumptions. While for this paper we have disregarded the rotation curve term (equation \ref{eqn:poisson}), we note that an accurate determination of this will be necessary for the implementation of this or other $z$-direction methods to real data.

Non-spherical DM density profiles, such as oblate halos or accreted dark discs, can also be fitted using our method by incorporating a dark disc term into the reconstruction model, which is shown in Fig.~\ref{fig:darkdisc1E6}. Our method can also reconstruct mock data sets containing both a tilt term and a dark disc, as shown in Fig.~\ref{fig:DD_tilt_1E6}.

\section{Acknowledgments} \nonumber
We thank Michael Feyereisen, Sebastian Liem, Pat Scott, Roberto Trotta, Stephen Feeney, Roberto Ruiz de Austri, Scott Tremaine, Vera Gluscevic, and Juna Kollmeier for very useful discussions. S.S. acknowledges support from the Swedish Research Council (VR, Contract No. 350-2012-268). J.I.R. would like to acknowledge support from STFC consolidated grant ST/M000990/1 and the MERAC foundation. G.B. acknowledges support from the European Research Council through the ERC starting grant \textit{WIMPs Kairos}.

\bibliographystyle{mnras}
\bibliography{LocalDM_Bibliography}

\appendix
\section{Variation due to the number of bins}\label{app:binning}
\begin{figure*}
\includegraphics[width=0.95\textwidth,clip]{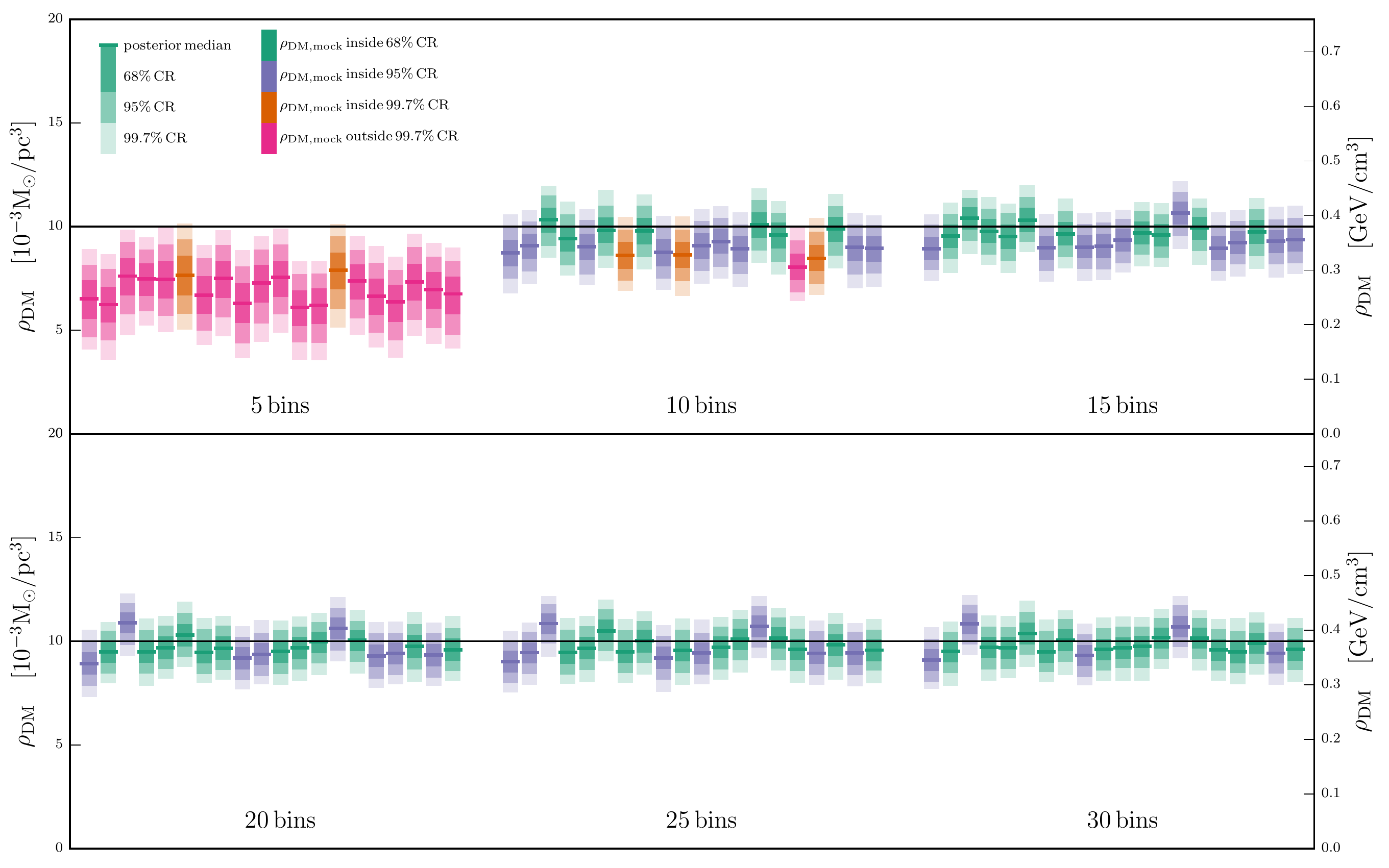}
\caption{Exploring the effects of binning on the determination of $\rho_{\rm DM}$. These plots show marginalised posteriors for $\rho_{\rm DM}(z) = \rho_{\rm DM, const}$ for the 20 mock data sets {\tt thick\_1E6\_0-19}, reconstructed using 5, 10, 15, 20, 25, and 30 bins. Dark, medium, and light shading show the 68\%, 95\%, and 99.7\% credible regions (CRs) respectively. Green, purple, and orange colouring indicates that the 68\%, 95\%, or 99.7\% CR respectively contains the correct answer, while pink colouring indicated that the correct answer lies outside even the 99.7\% CR.. The median value of each posterior is shown by a solid line in green, purple, or orange, while the DM density value used to generate the mock data is shown as a solid black line across the entire plot.}
\label{fig:binning_1E6}
\end{figure*}

Figure \ref{fig:binning_1E6} illustrates the effects of changing the number of bins used for this analysis. There we plot the 68\%, 95\%, or 99.7\% CRs for 20 mock data sets ({\tt thick\_1E6\_0-19}, no tilt and no DD), and vary the number of bins from five to 30, in increments of five. For only five bins (top left set), the true answer for $\rho_{\rm DM, const}$ is outside the 99.7\% CR for all but two of the mocks, for which the true answer is within only the 99.7\% CR. The systematic underestimation is due to an over estimation of the baryonic disk. The fifth bin in this scheme covers a range from 1.2 kpc to 2.4 kpc and has its bin centre at $z= 1.6$kpc, so it is unsurprising that such a low number of bins fails to correctly reconstruct the DM profile, which is a subdominant component only becomes apparent at higher $z$. Increasing the number of bins to 10, 15, and then to 20 improves the results. The gains from increasing from 20 to 25 and 30 bins is very slight. 

\section{Dark Disc Reconstructions with Multiple Mocks}\label{app:dd_grids}
Here we show the results of reconstructing all mocks data sets generated using a constant DM density only ({\tt thick\_X\_M}), constant DM and a regular dark disc ({\tt thick\_dd\_X\_M}), or constant DM and a `big' dark disc ({\tt thick\_bdd\_X\_M}); see Table \ref{tab:mockdataparams} for details. The reconstruction models either have a constant DM density only, or a constant DM density plus a dark disc.

\begin{figure*}
\includegraphics[width=0.95\textwidth,clip]{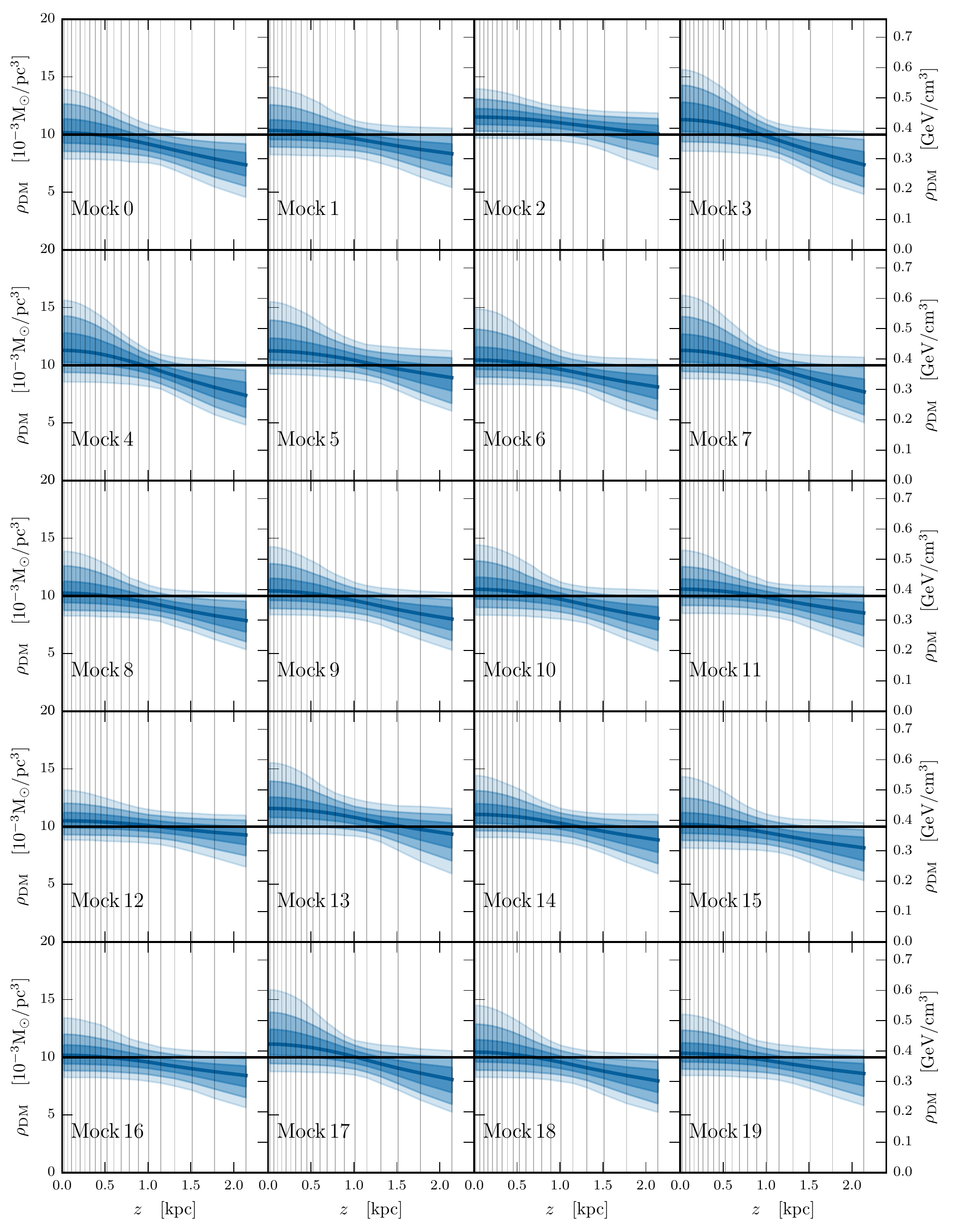}
\caption{Marginalized posteriors of $\rho_{\rm DM}(z)$ for mock data sets {\tt thick\_1E6\_0-19} with $\rho_{\rm DM, const}$ only (no DD component) reconstructed using a model with $\rho_{\rm DM, const}$ plus a DD. Dark, medium, and light shading indicate the 68\%, 95\%, and 99.7\% CRs respectively. The median value of the posterior is shown as the solid blue line, while the DM density profile used to generate the mock data is shown as a solid black line.}
\label{fig:grid_MockNoDD_ModelDD}
\end{figure*}

\begin{figure*}
\includegraphics[width=0.95\textwidth,clip]{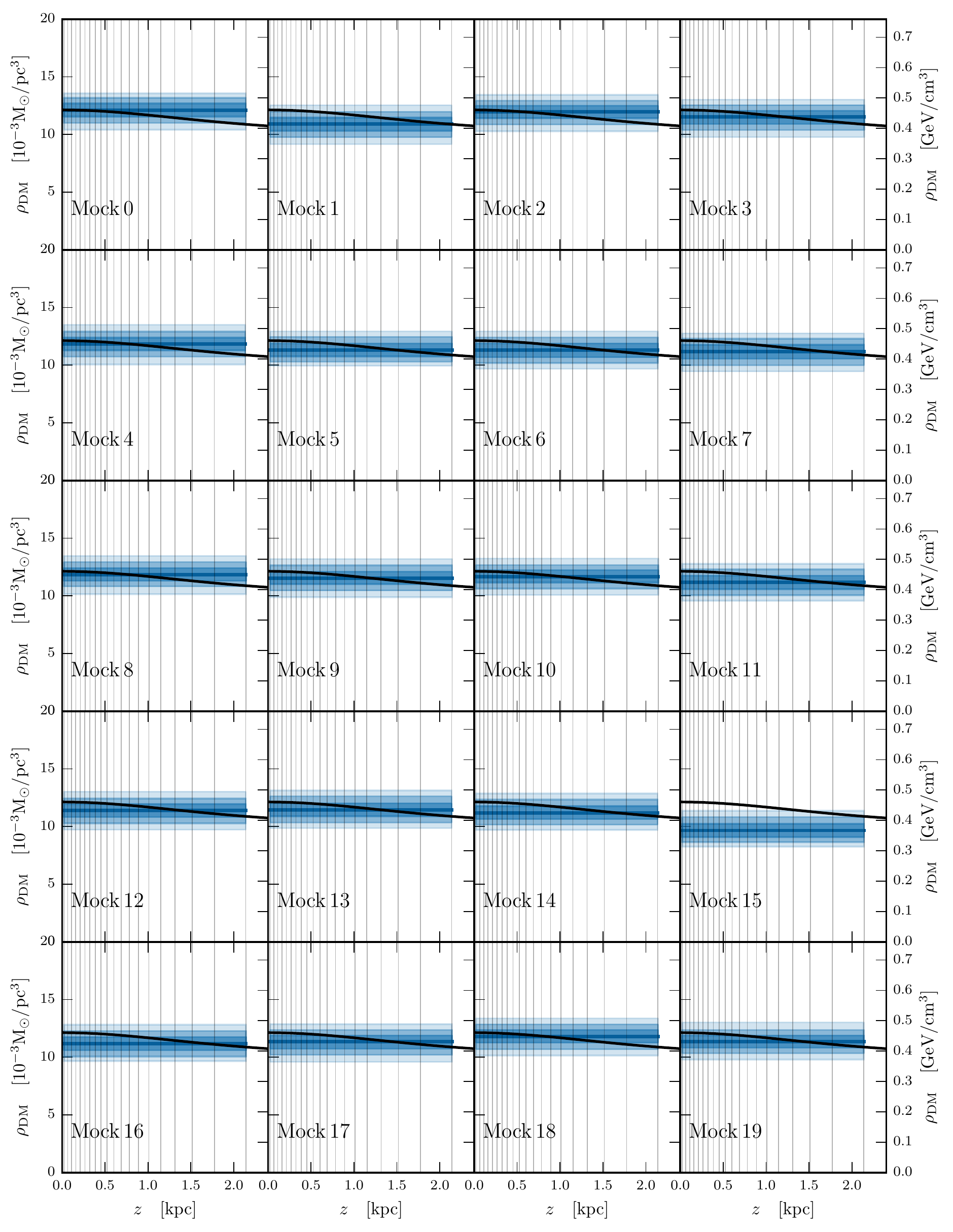}
\caption{Marginalized posteriors of $\rho_{\rm DM}(z)$ for mock data sets {\tt thick\_dd\_1E6\_0-19} with $\rho_{\rm DM, const}$ plus a DD component, reconstructed using a model with $\rho_{\rm DM, const}$ only. Dark, medium, and light shading indicate the 68\%, 95\%, and 99.7\% CRs respectively. The median value of the posterior is shown as the solid blue line, while the DM density profile used to generate the mock data is shown as a solid black line.}
\label{fig:grid_MockDD_ModelNoDD}
\end{figure*}

\begin{figure*}
\includegraphics[width=0.95\textwidth,clip]{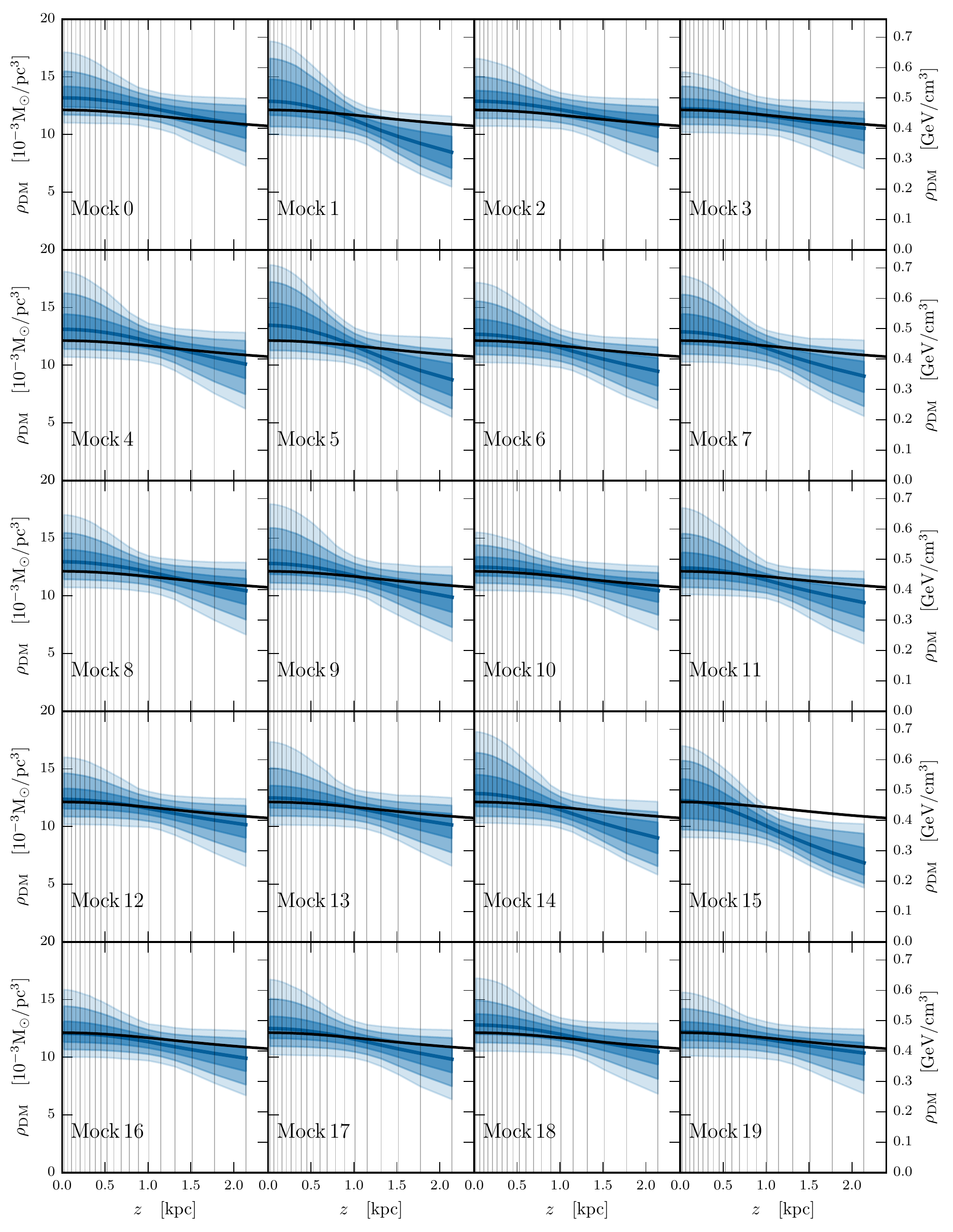}
\caption{Marginalized posteriors of $\rho_{\rm DM}(z)$ for mock data sets {\tt thick\_dd\_1E6\_0-19} with $\rho_{\rm DM, const}$ plus a DD component, reconstructed using a model with $\rho_{\rm DM, const}$ and a DD component. Dark, medium, and light shading indicate the 68\%, 95\%, and 99.7\% CRs respectively. The median value of the posterior is shown as the solid blue line, while the DM density profile used to generate the mock data is shown as a solid black line.}
\label{fig:grid_MockDD_ModelDD}
\end{figure*}

\begin{figure*}
\includegraphics[width=0.95\textwidth,clip]{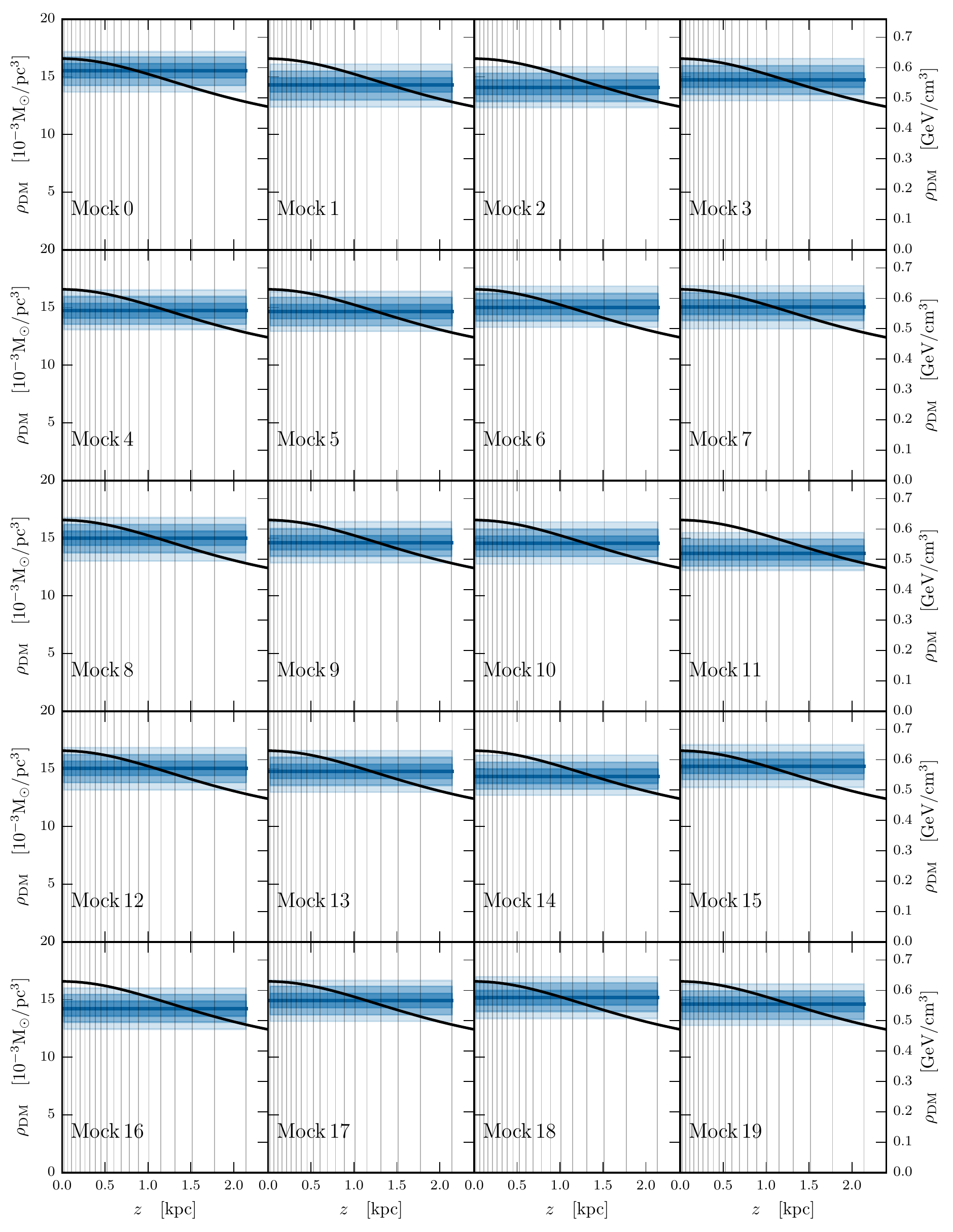}
\caption{Marginalized posteriors of $\rho_{\rm DM}(z)$ for mock data sets {\tt thick\_bdd\_1E6\_0-19} with $\rho_{\rm DM, const}$ plus a DD component, reconstructed using a model with $\rho_{\rm DM, const}$ only. Dark, medium, and light shading indicate the 68\%, 95\%, and 99.7\% CRs respectively. The median value of the posterior is shown as the solid blue line, while the DM density profile used to generate the mock data is shown as a solid black line.}
\label{fig:grid_MockBDD_ModelNoDD}
\end{figure*}

\begin{figure*}
\includegraphics[width=0.95\textwidth,clip]{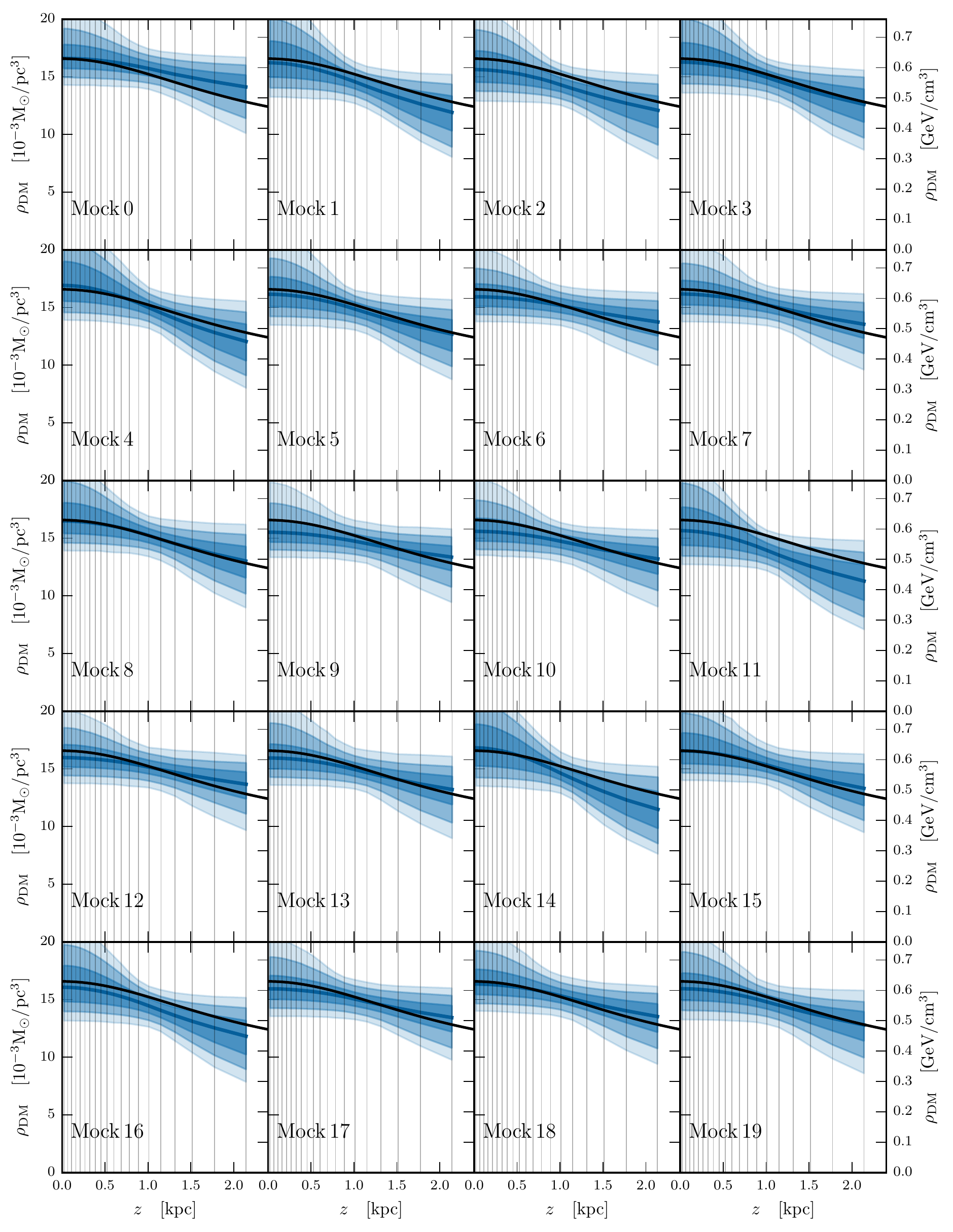}
\caption{Marginalized posteriors of $\rho_{\rm DM}(z)$ for mock data sets {\tt thick\_bdd\_1E6\_0-19} with $\rho_{\rm DM, const}$ plus a `big' DD component, reconstructed using a model with $\rho_{\rm DM, const}$ and a DD component. Dark, medium, and light shading indicate the 68\%, 95\%, and 99.7\% CRs respectively. The median value of the posterior is shown as the solid blue line, while the DM density profile used to generate the mock data is shown as a solid black line.}
\label{fig:grid_MockBDD_ModelNoDD}
\end{figure*}

\begin{figure*}
\includegraphics[width=0.95\textwidth,clip]{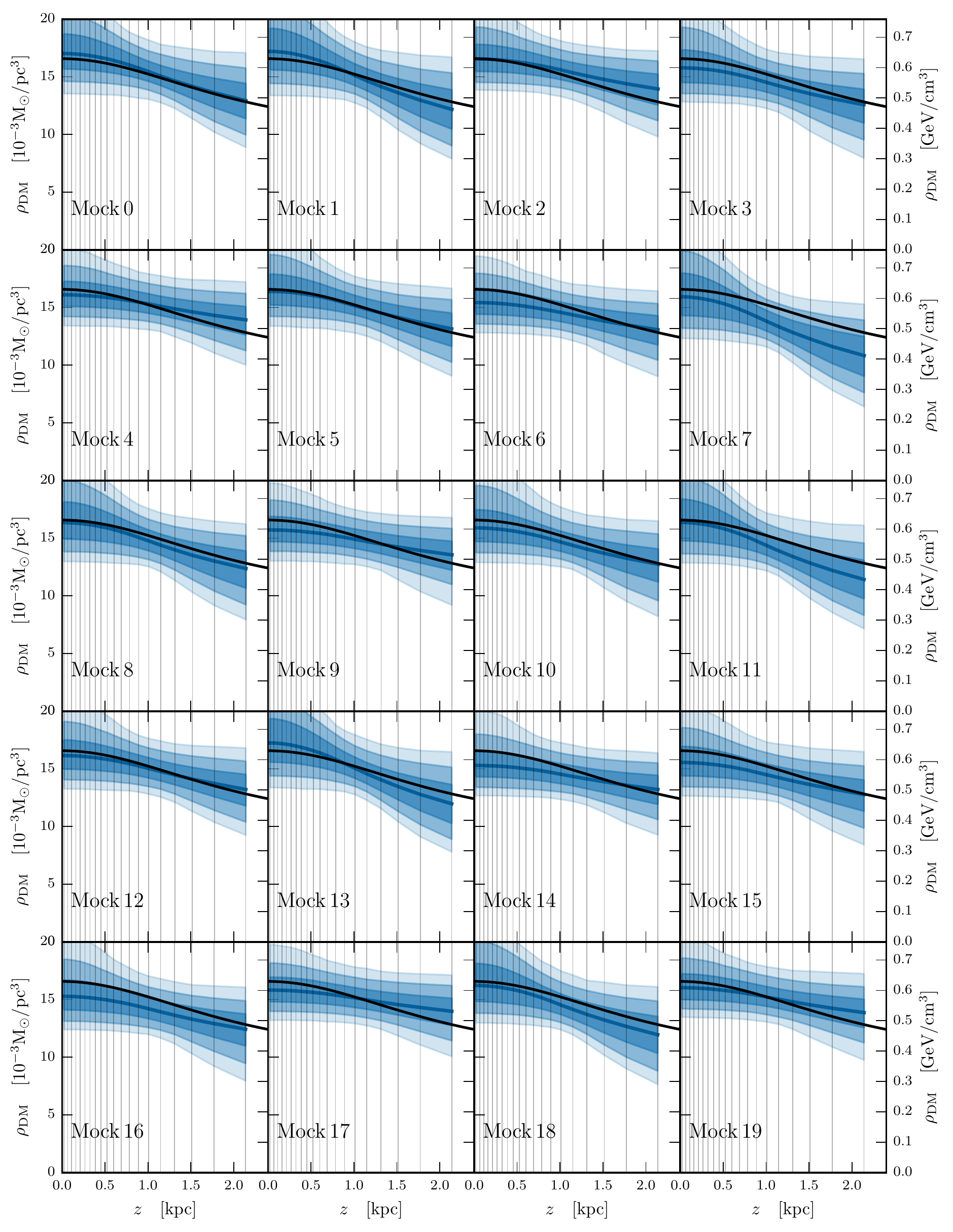}
\caption{Marginalized posteriors of $\rho_{\rm DM}(z)$ for mock data sets {\tt thick\_bdd\_tilt\_X\_0-19} with $\rho_{\rm DM, const}$ plus a `big' DD component and including the effects of tilt, reconstructed using a model with $\rho_{\rm DM, const}$, a DD component, and incorporating tilt. Dark, medium, and light shading indicate the 68\%, 95\%, and 99.7\% CRs respectively. The median value of the posterior is shown as the solid blue line, while the DM density profile used to generate the mock data is shown as a solid black line.}
\label{fig:grid_MockBDD_MockTilt_ModelTilt_ModelDD}
\end{figure*}

\bsp	
\label{lastpage}
\end{document}